\documentclass[11pt,a4paper]{article}
\pdfoutput=1
\usepackage{jheppub}
\usepackage{tikz}
\usetikzlibrary{intersections, calc, arrows.meta, bending}
\usepackage{pgfplots}
\usetikzlibrary{intersections, pgfplots.fillbetween}
\usepackage{orcidlink}
\usepackage{bbm}

\DeclareMathOperator{\Tr}{Tr}

\newcommand{\ri}{\mathrm{i}}
\renewcommand{\th}{\theta}

\newcommand{\cob}{\delta}

\newcommand{\hf}{\frac{1}{2}}
\newcommand{\qu}{\frac{1}{4}}
\newcommand{\til}[1]{\widetilde{#1}}
\newcommand{\si}{\sigma}

\renewcommand{\b}[1]{\overline{#1}}
\newcommand{\del}{\partial}

\newcommand{\bra}{\langle}
\newcommand{\ket}{\rangle}

\newcommand{\bt}{\beta}
\newcommand{\ga}{\gamma}

\newcommand{\al}{\alpha}

\newcommand{\rt}[1]{\sqrt{#1}}
\newcommand{\cO}{\mathcal{O}}
\newcommand{\cZ}{\mathcal{Z}}
\newcommand{\cF}{\mathcal{F}}

\newcommand{\cD}{\mathcal{D}}
\newcommand{\cH}{\mathcal{H}}

\newcommand{\cT}{\mathcal{T}}
\newcommand{\cN}{\mathcal{N}}

\newcommand{\cC}{\mathcal{C}}

\newcommand{\mat}[1]{\begin{pmatrix}#1\end{pmatrix}}
\newcommand{\bmat}[1]{\begin{bmatrix}#1\end{bmatrix}}

\newcommand{\id}{\mathbbm{1}}

\newcommand{\Zemp}{\tikz[baseline={([yshift=-.5ex]current bounding box.center)}]{
\draw[thick] (0,0)--(1,0)--(1,1)--(0,1)--(0,0);}}
\newcommand{\Zvp}{\tikz[baseline={([yshift=-.5ex]current bounding box.center)}]{
\draw[thick] (0,0)--(1,0)--(1,1)--(0,1)--(0,0);
\draw[very thick,red] (0.5,0)--(0.5,1);}}
\newcommand{\Zvs}{\tikz[baseline={([yshift=-.5ex]current bounding box.center)}]{
\draw[thick] (0,0)--(1,0)--(1,1)--(0,1)--(0,0);
\draw[very thick,green] (0.5,0)--(0.5,1);}}
\newcommand{\Zhp}{\tikz[baseline={([yshift=-.5ex]current bounding box.center)}]{
\draw[thick] (0,0)--(1,0)--(1,1)--(0,1)--(0,0);
\draw[very thick,red] (0,0.5)--(1,0.5);}}
\newcommand{\Zhs}{\tikz[baseline={([yshift=-.5ex]current bounding box.center)}]{
\draw[thick] (0,0)--(1,0)--(1,1)--(0,1)--(0,0);
\draw[very thick,green] (0,0.5)--(1,0.5);}}
\newcommand{\Ztp}{\tikz[baseline={([yshift=-.5ex]current bounding box.center)}]{
\draw[thick] (0,0)--(1,0)--(1,1)--(0,1)--(0,0);
\draw[very thick,red] (0,0.5) to [out=0, in=-90] (0.5,1);
\draw[very thick,red] (0.5,0) to [out=90, in=180] (1,0.5);
}}
\newcommand{\Zts}{\tikz[baseline={([yshift=-.5ex]current bounding box.center)}]{
\draw[thick] (0,0)--(1,0)--(1,1)--(0,1)--(0,0);
\draw[very thick,green] (0,0.5) to [out=0, in=-90] (0.5,1);
\draw[very thick,green] (0.5,0) to [out=90, in=180] (1,0.5);
}}

\newcommand{\Ztsr}{\tikz[baseline={([yshift=-.5ex]current bounding box.center)}]{
\draw[thick] (0,0)--(1,0)--(1,1)--(0,1)--(0,0);
\draw[very thick,green] (0,0.5) to [out=0, in=-90] (0.5,1);
\draw[very thick,green] (0.5,0) to [out=90, in=180] (1,0.5);
\draw[very thick, red] (0.35,0.65)--(0.65,0.35);
}}

\newcommand{\Zsp}{\tikz[baseline={([yshift=-.5ex]current bounding box.center)}]{
\draw[thick] (0,0)--(1,0)--(1,1)--(0,1)--(0,0);
\draw[very thick, green] (0.35,0.65)--(0.65,0.35);
\draw[very thick, green] (0.5,1) arc (0:-45:0.5);
\draw[very thick, red] (0,0.5) arc (-90:-45:0.5);
\draw[very thick, green] (0.5,0) arc (180:135:0.5);
\draw[very thick, red] (1,0.5) arc (90:135:0.5);
}}

\newcommand{\Zps}{\tikz[baseline={([yshift=-.5ex]current bounding box.center)}]{
\draw[thick] (0,0)--(1,0)--(1,1)--(0,1)--(0,0);
\draw[very thick, green] (0.35,0.65)--(0.65,0.35);
\draw[very thick, red] (0.5,1) arc (0:-45:0.5);
\draw[very thick, green] (0,0.5) arc (-90:-45:0.5);
\draw[very thick, red] (0.5,0) arc (180:135:0.5);
\draw[very thick, green] (1,0.5) arc (90:135:0.5);
}}

\makeatletter
\gdef\@fpheader{}
\makeatother

\begin{document}
\title{Holographic dual of the 2d Ising model}

\author{Kazumi Okuyama\,\orcidlink{0000-0002-3591-6954}\,}
\affiliation{Department of Physics, 
Shinshu University\\ 3-1-1 Asahi, Matsumoto 390-8621, Japan}

\emailAdd{ kazumi@azusa.shinshu-u.ac.jp}

\abstract{
We propose a holographic dual of the 2d Ising model on
a torus as the Turaev-Viro TQFT on a solid torus.
The bulk spacetime is divided into the ``core'' region
and the ``sandwich'' region.
The core region prepares a defect state $|\mathcal{D}\rangle$
on the topological boundary $\mathbb{D}$.
The sandwich region computes the Turaev-Viro state sum
$\langle\mathcal{D}|\Pi|\mathcal{F}\rangle$ 
between the state $|\mathcal{D}\rangle$ on 
$\mathbb{D}$ and the state $|\mathcal{F}\rangle$ on the physical
boundary $\mathbb{B}$.
The fusion diagram (or string-net) 
$\mathcal{F}$ on 
$\mathbb{B}$ is given by the Aasen-Fendley-Mong construction.
There is also a ``bedding surface'' $\mathbb{S}$
between $\mathbb{D}$ and $\mathbb{B}$, and
the height variables 
(or Ising spins inside/outside of the domain wall)
are naturally living on
$\mathbb{S}$.
The bulk spacetime is discretized by the 
2-skeleton $P$ which interpolates
$\mathcal{D}$ and $\mathcal{F}$.  
When the defect $\mathcal{D}$ is absent,
the 2-skeleton $P$ ends on 
the bedding surface $\mathbb{S}$ and $P$ is interpreted as the bulk worldsheet of the 
string-net $\mathcal{F}$.}

\maketitle

\section{Introduction}
The holographic duality and the
AdS/CFT correspondence are the main
subject in the study of quantum gravity \cite{Witten:1998qj}.
The most famous example is the duality
between 4d $\cN=4$ super Yang-Mills theory
and the type IIB string theory on $AdS_5\times S^5$
\cite{Maldacena:1997re}.
However, in general 
it is not so
easy to find a bulk dual for a given boundary theory.

Recently, the holographic duality
between the 2d JT gravity 
\cite{Jackiw:1984je,Teitelboim:1983ux} and the SYK model \cite{Sachdev1993,Kitaev1,Kitaev2}
has attracted a lot of attention.
Furthermore, this duality is understood as a certain limit
of the duality between the 2d sine-dilaton gravity 
\cite{Blommaert:2024ymv} and the
double-scaled SYK (DSSYK) model \cite{Berkooz:2018jqr}.

Interestingly, in \cite{Okuyama:2026aoa}
it was observed that
the transfer matrix of DSSYK also appears in 
a completely different model called 
ASEP (asymmetric simple exclusion process) \cite{derrida1993exact}.
It was further observed 
in \cite{Okuyama:2026aoa} that this relation
between ASEP and DSSYK has an analogue in the 
3d holography between the Turaev-Viro TQFT 
\cite{turaev1992state} and
the Levin-Wen string-net model \cite{Levin:2004mi}. 

In this paper, we will make the observation 
in \cite{Okuyama:2026aoa} more concrete.
We construct a holographic dual of the 2d Ising
model on a torus as the Turaev-Viro TQFT on a solid torus.
In \cite{Barrett:1993ab}, 
Barrett and Westbury generalized the 
Turaev-Viro state sum to 
the state sum for an 
arbitrary fusion category $\cC$.
To find a holographic dual of the Ising model,
we first rewrite the 
low temperature expansion of the Ising model
as a trivalent fusion diagram $\cF$
using the construction of Aasen-Fendley-Mong (AFM) 
\cite{Aasen:2020jwb}.
The AFM construction allows us to write a trivalent
fusion diagram even on a square lattice 
(see \eqref{eq:fusion-lattice}). 

We propose 
that the bulk dual of the 2d Ising model on
a torus is the Turaev-Viro TQFT on a solid torus.
As shown in \eqref{eq:cross-section},  
the bulk spacetime is divided into the ``core'' region
and the ``sandwich'' region
between the topological boundary $\mathbb{D}$
and the physical boundary $\mathbb{B}$.
The ``core'' region can be thought of as the state
preparation of the defect state $|\cD\ket$
on $\mathbb{D}$ (see \eqref{eq:map-mu-D}).
The ``sandwich'' region computes
the Turaev-Viro state sum $\bra\cD|\Pi|\cF\ket$
between the defect state $|\cD\ket$ on $\mathbb{D}$
and the state $|\cF\ket$ on $\mathbb{B}$.
There is also a ``bedding surface'' $\mathbb{S}$ between
the topological boundary $\mathbb{D}$
and the physical boundary $\mathbb{B}$ 
(see \eqref{eq:cross-section}).
In the shadow world construction of Turaev \cite{turaev1992shadow},
the height variables naturally live on the bedding surface
$\mathbb{S}$; the 
height variables for the Ising model
correspond to the spins inside/outside of the domain wall
in the low temperature expansion (see \eqref{eq:S-conf1} and \eqref{eq:S-conf2}).
Following the approach in \cite{turaev2013,turaev2017monoidal},
we discretize the 
bulk spacetime by the 
2-skeleton $P$ and we assign ``colors'' $c\in\cC$
to the faces of $P$.
The 2-skeleton $P$ interpolates
$\mathcal{D}$ and $\mathcal{F}$
in such a way that the coloring of the faces of $P$
naturally extends to the coloring of 
$\del P=\cD\cup\cF$.  
When the defect $\mathcal{D}$ is absent,
the 2-skeleton $P$ ends on 
the bedding surface $\mathbb{S}$ and $P$ is interpreted as the bulk worldsheet of the 
string-net $\mathcal{F}$ (see \eqref{eq:P-bulk}).

This paper is organized as follows. In section \ref{sec:review},
we review the high and the low temperature expansions of 
the 2d Ising model.
In section \ref{sec:AFM}, we introduce the 
AFM construction of the fusion diagram
for the Ising model in the low temperature expansion.
In section \ref{sec:CFT}, we consider the
torus partition functions of Ising CFT with defects. 
In section \ref{sec:tube},
we compute the minimal central idempotents in the tube algebra
of
the Ising category.
In section \ref{sec:holography}, we construct the bulk dual
of the 2d Ising model on a torus
as the Turaev-Viro TQFT on a solid torus.
Finally, we conclude in section
\ref{sec:conclusion} with some discussions on the future problems.
In appendix \ref{app:SL2}, we consider the sum over the $SL(2,\mathbb{Z})$
images of the vacuum character $|\chi_\id|^2$ of Ising CFT.

\section{Review of Ising model}\label{sec:review}
We consider the 2d Ising model on a square lattice
with a ferromagnetic interaction $J>0$
\begin{equation}
\begin{aligned}
 Z_\text{Ising}=
\sum_{\{s_v=\pm1\}}\exp\left(J\sum_{\bra v,v'\ket} s_vs_{v'}\right).
\end{aligned} 
\label{eq:Ising-def}
\end{equation}
Here $v$ runs over the vertex of the square lattice and
$\bra v,v'\ket$ denotes the nearest neighbors.
For simplicity, we consider the homogeneous model, where the coupling $J$
in the horizontal direction and the vertical direction are equal.  
Our model is defined on 
a $L_1\times L_2$ square lattice,
and we sometimes denote $v=(i,j)$ with
\begin{equation}
\begin{aligned}
 i=0,\cdots, L_1-1,\quad
j=0,\cdots, L_2-1.
\end{aligned} 
\end{equation}
The number of sites and links
are given by
\begin{equation}
\begin{aligned}
 \#\text{site}=L_1L_2,\quad \#\text{link}=2L_1L_2.
\end{aligned} 
\end{equation}
We also impose the periodic boundary conditions on both 
directions $i$ and $j$
\begin{equation}
\begin{aligned}
 s_{i+L_1,j}=s_{i,j+L_2}=s_{i,j}.
\end{aligned} 
\end{equation}
Namely, we study the 2d Ising model on a torus.

It is well-known that the 
$Z_\text{Ising}$ in \eqref{eq:Ising-def}
admits the high and the low temperature
expansions and these expansions are exchanged by
the Kramers-Wannier duality \cite{kramers1941,kramers1941-2} 
(see the textbook of Itzykson-Drouffe \cite{itzykson1989statistical}
for a standard review).

Let us first consider the high temperature expansion $J\ll1$.
Using the identity
\begin{equation}
\begin{aligned}
 e^{Js_vs_{v'}}=\cosh J\times (1+ts_vs_{v'}),\qquad t=\tanh J,
\end{aligned} 
\label{eq:def-t}
\end{equation}
$Z_\text{Ising}$ in \eqref{eq:Ising-def} is written as
\begin{equation}
\begin{aligned}
 Z_\text{Ising}=(\cosh J)^{\#\text{link}}
\sum_{\{s_v=\pm1\}}\prod_{\bra v,v'\ket}(1+ts_vs_{v'}).
\end{aligned} 
\label{eq:high-sum}
\end{equation}
Expanding the summand of \eqref{eq:high-sum}
in the small $t$ regime and using the relation, 
\begin{equation}
\begin{aligned}
 \sum_{s_v=\pm1}(s_v)^{n}=2\cob_{n\,(\text{mod}\,2),0}
\end{aligned} 
\end{equation}
we arrive at the loop expansion of $Z_\text{Ising}$
\begin{equation}
\begin{aligned}
 Z_\text{Ising}=2^{\#\text{site}}
(\cosh J)^{\#\text{link}}\sum_{\text{closed~loop}~\ga} t^{|\ga|},
\end{aligned} 
\label{eq:high-loop}
\end{equation}
where $|\ga|$ denotes the length of the loop $\ga$.

On the other hand, in the low temperature regime $J\gg1$
the 2d lattice is divided into
the domain of up spins $s_v=+1$  
and the domain of down spins $s_v=-1$.
The boundary of two types of domains, 
called the domain wall, costs the factor of $e^{-2J}$ for each edge of the wall 
due to the relation
\begin{equation}
\begin{aligned}
 e^{Js_vs_{v'}}=e^{J}\times(\cob_{s_vs_{v'},1}+e^{-2J}\cob_{s_vs_{v'},-1}).
\end{aligned} 
\end{equation}
Thus we find the low temperature expansion
\begin{equation}
\begin{aligned}
 Z_\text{Ising}=e^{(\#\text{link})J}\sum_{\text{closed~loop}~\ga} (e^{-2J})^{|\ga|}.
\label{eq:low-loop}
\end{aligned} 
\end{equation}
The two expansions \eqref{eq:high-loop} and
\eqref{eq:low-loop} are exchanged by
introducing the dual coupling $\til{J}$
\begin{equation}
\begin{aligned}
 \tanh J=e^{-2\til{J}}.
\end{aligned} 
\label{eq:J-tilJ}
\end{equation}
We can invert this relation \eqref{eq:J-tilJ}
and express $\tanh\til{J}$ in terms of $J$.
It turns out that the inverse relation takes the same form as
\eqref{eq:J-tilJ}
\begin{equation}
\begin{aligned}
 \tanh \til{J}=e^{-2J}.
\end{aligned} 
\label{eq:tanhtilJ}
\end{equation}
One can also show the following relation 
which is symmetric in $J$ and $\til{J}$
\begin{equation}
\begin{aligned}
 \sinh (2J)\sinh(2\til{J})=1.
\end{aligned} 
\label{eq:JtilJ-sym}
\end{equation}

It is also convenient to introduce the transfer matrix $T$
which acts on the state of spins along the horizontal direction
with a fixed $j$
\begin{equation}
\begin{aligned}
 |s_{0,j}\ket\otimes |s_{1,j}\ket\otimes \cdots
\otimes |s_{L_1-1,j}\ket \in \cH_{\text{spin}},
\end{aligned} 
\end{equation}
where $\cH_{\text{spin}}$ denotes the Hilbert space
of spins at a fixed $j$
\begin{equation}
\begin{aligned}
 \cH_\text{spin}=\bigotimes_{i=0}^{L_1-1} \mathbb{C}^2_i,\qquad 
\mathbb{C}^2_i=\mathbb{C}|+\ket_i+\mathbb{C}{|-\ket}_i.
\end{aligned} 
\label{eq:H-spin}
\end{equation}

Using the transfer matrix $T$, the partition function
is written as
\begin{equation}
\begin{aligned}
 Z_\text{Ising}=\Tr T^{L_2}.
\end{aligned} 
\label{eq:Z-trace}
\end{equation}
The transfer matrix $T$ is decomposed into the horizontal part $T_H$ and
the vertical part $T_V$
\begin{equation}
\begin{aligned}
 T=T_VT_H.
\end{aligned} 
\end{equation}
If the link $\bra v,v'\ket$ is along the horizontal direction,
i.e. $v=(i,j)$ and $v'=(i+1,j)$,
the interaction $e^{Js_vs_{v'}}$ is diagonal in the spin basis 
$|\pm\ket_i\otimes |\pm\ket_{i+1}$ in
\eqref{eq:H-spin}.
Thus $T_H$ is written in terms of 
the Pauli $\si^z$-matrix acting on the factor $\mathbb{C}^2_i$
at site $i$  
\begin{equation}
\begin{aligned}
 \si_i^z|\pm\ket_i=\pm|\pm\ket_i.
\end{aligned} 
\end{equation}
In this way we find
\begin{equation}
\begin{aligned}
 T_H=\exp\left(J\sum_{i=0}^{L_1-1}\si_i^z\si_{i+1}^z\right).
\end{aligned} 
\label{eq:TH}
\end{equation}
On the other hand, if the link
$\bra v,v'\ket$ is along the vertical direction
the interaction $e^{Js_vs_{v'}}$ is not diagonal in the spin basis $|\pm\ket_i$.
If we identify $s_v$ at $v=(i,j)$ as ket $|s_v\ket_i$ and $s_{v'}$ 
at $v'=(i,j+1)$ as bra ${}_i\bra s_{v'}|$,
the interaction $e^{Js_vs_{v'}}$ is written as a $2\times 2$ matrix
\begin{equation}
\begin{aligned}
 \bigl(e^{Js_vs_{v'}}\bigr)_{s_{v'},s_v=\pm}=
\mat{e^{J} ~&e^{-J} \\ e^{-J}~& e^{J}}=e^{J}\bigl(1+e^{-2J}\si^x_i\bigr).
\end{aligned} 
\end{equation}
Using the relation \eqref{eq:tanhtilJ}, this matrix is written as
\begin{equation}
\begin{aligned}
e^{J}\bigl( 1+e^{-2J}\si^x_i\bigr)&=\frac{e^J}{\cosh\til{J}}\,\bigl(\cosh\til{J}+\si^x_i\sinh\til{J}\bigr)\\
&=\frac{e^J}{\cosh\til{J}} \,e^{\til{J}\si_i^x}
=\bigl(2\sinh 2J\bigr)^{\hf}\,e^{\til{J}\si_i^x}.
\end{aligned} 
\end{equation}
Thus the vertical part $T_V$ of the transfer matrix is given by
\begin{equation}
\begin{aligned}
 T_V=\prod_{i=0}^{L_1-1}\bigl(2\sinh 2J\bigr)^{\hf}e^{\til{J}\si_i^x}\equiv
\bigl(2\sinh 2J\bigr)^{\hf L_1}T_V',
\end{aligned} 
\end{equation}
where $T_V'$ is given by
\begin{equation}
\begin{aligned}
 T_V'=\exp\left(\til{J}\sum_{i=0}^{L_1-1}\si_i^x\right).
\end{aligned} 
\end{equation}
Then the partition function \eqref{eq:Z-trace} becomes
\begin{equation}
\begin{aligned}
 Z_\text{Ising}=\bigl(2\sinh 2J\bigr)^{\hf L_1L_2}Z_\text{Ising}',\quad
Z_\text{Ising}'=\Tr (T_V'T_H)^{L_2}.
\end{aligned}
\label{eq:Z-factor} 
\end{equation}
It turns out that $Z_\text{Ising}'$ is invariant under the duality
$J\leftrightarrow \til{J}$.
This can be seen by introducing the disorder operator
$\mu_i^z$ via the Jordan-Wigner transformation
\begin{equation}
\begin{aligned}
 \mu_i^z=\prod_{i'=0}^{i-1}\si_{i'}^x.
\end{aligned} 
\end{equation}
Since $\si_i^x=\mu_{i}^z\mu_{i+1}^z$,  $T_V'$ is written as
\begin{equation}
\begin{aligned}
 T_V'=\exp\left(\til{J}\sum_{i=0}^{L_1-1}\mu_{i}^z\mu_{i+1}^z\right).
\end{aligned} 
\label{eq:TV'}
\end{equation}
From \eqref{eq:TH} and \eqref{eq:TV'}, it is clear that
$T_H$ and $T_V'$ are exchanged under the duality
$(J,\si_i^z)\leftrightarrow (\til{J},\mu_i^z)$.

The free energy per site is defined 
in the thermodynamic limit $L_1,L_2\to\infty$ as
\begin{equation}
\begin{aligned}
 F=\lim_{L_1,L_2\to\infty}\frac{1}{L_1L_2}\log Z_\text{Ising}.
\end{aligned} 
\end{equation}
From \eqref{eq:Z-factor}, the free energy 
per site is written as
\begin{equation}
\begin{aligned}
 F=\hf\log(2\sinh 2J)+F',
\end{aligned} 
\label{eq:F-F'}
\end{equation} 
where $F'$ is the free energy coming from $Z_\text{Ising}'$.
The explicit form of $F'$ is given by
the Onsager's celebrated exact result \cite{onsager1944}
\footnote{The free energy per site is easily evaluated by the technique of the
Grassmann integral
\cite{samuel1980use,itzykson1982ising,giuliani2004universality}
\begin{equation}
\begin{aligned}
 F=\log\bigl(2\cosh^2J\bigr)+\hf\int\frac{d^2p}{(2\pi)^2}\log\Bigl[(1+t^2)^2-2t(1-t^2)
(\cos p_1+\cos p_2)\Bigr].
\end{aligned} 
\end{equation}
We can rewrite this into the form \eqref{eq:onsager}, where
$m_\psi^2$ is given by
\begin{equation}
\begin{aligned}
 m_\psi^2=\frac{(1+t^2)^2}{t(1-t^2)}-4=\left(2\rt{\frac{t}{1-t^2}}-\rt{\frac{1-t^2}{t}}\right)^2.
\end{aligned} 
\end{equation}
One can easily show that this agrees with $m_\psi$ in \eqref{eq:m-psi}.
The constant piece in \eqref{eq:F-F'} is also correctly reproduced
\begin{equation}
\begin{aligned}
 \log\bigl(2\cosh^2J\bigr)+\hf\log\bigl[t(1-t^2)\bigr]=\hf\log\bigl(2\sinh 2J\bigr).
\end{aligned} 
\end{equation}
}
\begin{equation}
\begin{aligned}
 F'
=\hf\int\frac{d^2p}{(2\pi)^2}\log\left[\left(2\sin\frac{p_1}{2}\right)^2+\left(2\sin\frac{p_2}{2}\right)^2+m_\psi^2\right].
\end{aligned} 
\label{eq:onsager}
\end{equation}
Here the integration region is $p_1,p_2\in[-\pi,\pi]$ and $m_\psi$ 
is the fermion mass
\begin{equation}
\begin{aligned}
 m_\psi=\bigl(2\sinh 2J\bigr)^\hf-\bigl(2\sinh2\til{J}\bigr)^\hf.
\end{aligned}
\label{eq:m-psi} 
\end{equation}
Note that $m_\psi$ changes sign under the duality $J\leftrightarrow\til{J}$,
while $F'$ 
in \eqref{eq:onsager} is duality invariant as we explained above.

At the critical point
\begin{equation}
\begin{aligned}
 J=\til{J}=J_c,
\end{aligned} 
\end{equation}
the fermion mass $m_\psi$ vanishes and the 
Ising model becomes the $c=1/2$ CFT in the IR.
From \eqref{eq:JtilJ-sym}, the critical value $J_c$ satisfies
\begin{equation}
\begin{aligned}
 \sinh2J_c=1,
\end{aligned} 
\end{equation}
which is solved as 
\begin{equation}
\begin{aligned}
 e^{-2J_c}=\rt{2}-1.
\end{aligned} 
\end{equation}

Note that the closed loop $\ga$ in the low temperature expansion 
\eqref{eq:low-loop} is on the dual lattice, while the loop $\ga$ in 
the high temperature expansion \eqref{eq:high-loop} is on the original lattice.
The loop $\ga$ in the low temperature expansion \eqref{eq:low-loop}
corresponds to the domain wall of spins. It turns out that the domain wall
can be thought of as the worldline of the Ising fermion $\psi$.
Note that the spins change sign as we cross the domain wall.
In other words, the worldline of $\psi$ creates a spin-flip defect.

\section{Height model and AFM construction}\label{sec:AFM}
In this section, we consider the fusion diagram expansion of the Ising model.
We follow the approach of \cite{Aasen:2020jwb}, which we will 
call the Aasen-Fendley-Mong
(AFM) construction. 

\subsection{Ising category}
Before going into the details of the AFM construction, let us summarize
the Ising fusion category, which we will utilize heavily in the rest of this paper.
In general, the fusion category $\cC$
is defined by the set of simple objects $\text{Irr}(\cC)$ obeying the fusion rule
\begin{equation}
\begin{aligned}
 a\otimes b=\sum_{c\in\text{Irr}(\cC)}N_{ab}^c\, c.
\end{aligned} 
\label{eq:fusion-N}
\end{equation}  
For the Ising model, the category $\cC$ consists of three simple objects
$\{\mathbbm{1},\psi,\si\}$ and the non-trivial fusion rules are given by
\begin{equation}
\begin{aligned}
 \psi\otimes\psi=\mathbbm{1},\quad
\si\otimes\si=\mathbbm{1}\oplus\psi,\quad
\psi\otimes\si=\si,
\end{aligned} 
\label{eq:fusion-Ising}
\end{equation}
and their quantum dimensions are given by
\begin{equation}
\begin{aligned}
 d_{\mathbbm{1}}=d_\psi=1,\quad d_\si=\rt{2}.
\end{aligned} 
\end{equation}
Note that the Ising category is a special case of the Tambara--Yamagami category \cite{tambara2000}.

We can draw a trivalent diagram 
respecting the fusion rule \eqref{eq:fusion-N}; such a diagram is called
the fusion diagram.
We also need the so-called $F$-symbol 
$[F^{abc}_d]_{x,y}$ for the 
$F$-move of fusion diagram
\begin{equation}
\begin{aligned}
 \begin{tikzpicture}[scale=1]
\draw[thick] (0,0)--(0.6,0);
\draw[thick] (0,0)--(-0.5,0.6);
\draw[thick] (0,0)--(-0.5,-0.6);
\draw[thick] (0.6,0)--(1.1,0.6);
\draw[thick] (0.6,0)--(1.1,-0.6);
\draw (0.3,0) node [below]{$x$};
\draw (-0.5,-0.7) node [below]{$a$};
\draw (-0.5,0.6) node [above]{$b$};
\draw (1.1,-0.6) node [below]{$d$};
\draw (1.1,0.6) node [above]{$c$};
\draw (1.3,0) node [right]{$\displaystyle =~\sum_{y} \bigl[F^{abc}_d\bigr]_{x,y}$};
\draw[thick] (4,0.6)--(4.3,0.25);
\draw[thick] (4.6,0.6)--(4.3,0.25);
\draw[thick] (4.3,0.25)--(4.3,-0.25);
\draw[thick] (4.6,-0.6)--(4.3,-0.25);
\draw[thick] (4,-0.6)--(4.3,-0.25);
\draw (4,0.6) node [above]{$b$};
\draw (4.6,0.6) node [above]{$c$};
\draw (4,-0.7) node [below]{$a$};
\draw (4.6,-0.6) node [below]{$d$};
\draw (4.3,0) node [right]{$y$};
\end{tikzpicture}
\end{aligned} ~.
\label{eq:F}
\end{equation}
For the Ising model, we have (see (4.21) in \cite{Aasen:2016dop})
\begin{equation}
\begin{aligned}
 \bigl[F_\psi^{\psi\psi\psi}\bigr]_{\mathbbm{1}\mathbbm{1}}=1,\quad
\bigl[F_\si^{\si\si\si}\bigr]_{ab}=\frac{(-1)^{ab}}{\rt{2}},\quad
\bigl[F_a^{\si b\si}\bigr]_{\si\si}=
\bigl[F_\si^{a\si b}\bigr]_{\si\si}=(-1)^{ab},
\end{aligned} 
\label{eq:F-Ising}
\end{equation}
where $a,b=0,1$ correspond to $\mathbbm{1},\psi$ respectively.
It is also convenient to introduce the so-called $6j$-symbol 
(or the tetrahedral symbol) by
\begin{equation}
\begin{aligned}
 \bmat{a~&b~&x\\ c~&d~&y}=\frac{1}{\rt{d_xd_y}}\bigl[F_d^{abc}\bigr]_{x,y}.
\end{aligned} 
\end{equation}
For $a,b,\chi=0,1$ we have
\begin{equation}
\begin{aligned}
 \bmat{\si~&\si~&a\\\si~&\si~&b}=\frac{(-1)^{ab}}{\rt{2}},\quad
\bmat{\si~&\si~&\chi\\a~&b~&\si}=N_{ab}^\chi=\cob_{|a-b|,\chi}.
\end{aligned} 
\end{equation}
Also, we will use the following relations in what follows:
\begin{equation}
\begin{aligned}
 \tikz[baseline={([yshift=-.5ex]current bounding box.center)},scale=1.6]{
\draw[thick] (0,0) circle (0.3); 
\node at (-0.3,0) [left] {$a$};
}~=~d_a,\qquad
 \tikz[baseline={([yshift=-.5ex]current bounding box.center)},scale=1.6]{
\draw[thick] (0,0) circle (0.2); 
\draw[thick] (0,0.2) -- (0,0.5);
\draw[thick] (0,-0.2) -- (0,-0.5);
\node at (-0.2,0) [left] {$c$};
\node at (0.17,0.03) [right] {$c'$};
\node at (0,0.5) [above] {$b$};
\node at (0,-0.5) [below] {$a$};
}~=~\cob_{a,b}\rt{\frac{d_cd_{c'}}{d_a}}
\tikz[baseline={([yshift=-.5ex]current bounding box.center)},scale=1.6]{
\draw[thick] (0,0.5) -- (0,-0.5);
\node at (0,0.5) [above] {$a$};
\node at (0,-0.5) [below] {$a$};
},
\end{aligned}
\label{eq:removal}
\end{equation}
\begin{equation}
\begin{aligned}
 \tikz[baseline={([yshift=-.5ex]current bounding box.center)},scale=1.6]{
\draw[thick] (0,-0.5) -- (0,0.5);
\draw[thick] (0.4,-0.5) -- (0.4,0.5);
\node at (0,0.5) [above] {$a$};
\node at (0,-0.8) [above] {$a$};
\node at (0.4,0.5) [above] {$b$};
\node at (0.4,-0.8) [above] {$b$};
}~=~\sum_{c\in\text{Irr}(\cC)}\rt{\frac{d_c}{d_ad_b}}~
\tikz[baseline={([yshift=-.5ex]current bounding box.center)},scale=1.6]{
\draw[thick] (0,0.4) arc (180:360:0.2);
\draw[thick] (0,-0.4) arc (180:0:0.2);
\draw[thick] (0.2,-0.2) -- (0.2,0.2);
\draw[thick] (0,0.4) -- (0,0.5);
\draw[thick] (0.4,0.4) -- (0.4,0.5);
\draw[thick] (0,-0.4) -- (0,-0.5);
\draw[thick] (0.4,-0.4) -- (0.4,-0.5);
\node at (0,0.5) [above] {$a$};
\node at (0,-0.8) [above] {$a$};
\node at (0.4,0.5) [above] {$b$};
\node at (0.4,-0.8) [above] {$b$};
\node at (0.2,0) [right] {$c$};
}~~.
\end{aligned}
\label{eq:cup-cap}
\end{equation}

\subsection{Height model}
Now, let us consider the fusion diagram expansion of the Ising model.
Naively, one might think that
the low temperature expansion \eqref{eq:low-loop}
looks like a fusion diagram where the domain wall corresponds to
a $\psi$-loop.
However, in order to make the diagram trivalent, we need to add small $\si$-circles
around the vertices of the dual lattice \cite{Aasen:2020jwb}.
This AFM construction allows us to write down the 
Boltzmann weight in a local form. 
As we will see below, the AFM construction
naturally gives rise to a height model, which is closely
related to the shadow world construction of Turaev \cite{turaev1992shadow}.

In the AFM construction \cite{Aasen:2020jwb}, we consider 
the bipartite lattice
\begin{equation}
\begin{aligned}
 \tikz[baseline={([yshift=-.5ex]current bounding box.center)}]{
\coordinate (a0) at (0,0); 
\coordinate (a1) at (1,0); 
\coordinate (a2) at (2,0); 
\coordinate (a3) at (3,0); 
\coordinate (s0) at (0.5,0.5);
\coordinate (s1) at (1.5,0.5); 
\coordinate (s2) at (2.5,0.5); 
\coordinate (s3) at (3.5,0.5); 
\coordinate (b0) at (0,1); 
\coordinate (b1) at (1,1); 
\coordinate (b2) at (2,1); 
\coordinate (b3) at (3,1); 
\coordinate (t0) at (0.5,1.5);
\coordinate (t1) at (1.5,1.5); 
\coordinate (t2) at (2.5,1.5); 
\coordinate (t3) at (3.5,1.5); 
\coordinate (c0) at (0,2); 
\coordinate (c1) at (1,2); 
\coordinate (c2) at (2,2); 
\coordinate (c3) at (3,2); 
\coordinate (u0) at (0.5,2.5);
\coordinate (u1) at (1.5,2.5); 
\coordinate (u2) at (2.5,2.5); 
\coordinate (u3) at (3.5,2.5);
\draw[dashed] (a0)--(s0)--(a1)--(s1)--(a2)--(s2)--(a3)--(s3);
\draw[dashed] (b0)--(s0)--(b1)--(s1)--(b2)--(s2)--(b3)--(s3);
\draw[dashed] (b0)--(t0)--(b1)--(t1)--(b2)--(t2)--(b3)--(t3);
\draw[dashed] (c0)--(t0)--(c1)--(t1)--(c2)--(t2)--(c3)--(t3);
\draw[dashed] (c0)--(u0)--(c1)--(u1)--(c2)--(u2)--(c3)--(u3);
\draw[green,thick] 
(s0) circle (0.2) (s1) circle (0.2) (s2) circle (0.2) (s3) circle (0.2)
(t0) circle (0.2) (t1) circle (0.2) (t2) circle (0.2) (t3) circle (0.2)
(u0) circle (0.2) (u1) circle (0.2) (u2) circle (0.2) (u3) circle (0.2); 
\fill 
(a0) circle (0.07) (a1) circle (0.07) (a2) circle (0.07) (a3) circle (0.07) 
(b0) circle (0.07) (b1) circle (0.07) (b2) circle (0.07) (b3) circle (0.07) 
(c0) circle (0.07) (c1) circle (0.07) (c2) circle (0.07) (c3) circle (0.07);
}
\end{aligned} 
\label{eq:lattice}
\end{equation}
where the green circles represent $\si$-loops while the black dots represent 
the physical spins $s_v=\pm1$.
Note that the black dots are on the original lattice, while the green circles
surround the vertices of the dual lattice.
It is convenient to introduce the height variable
$a=0,1$, which is related to the spin $s_v$ as
\begin{equation}
\begin{aligned}
 s_v=(-1)^a.
\end{aligned} 
\label{eq:sv-a}
\end{equation}

Now, let us consider the low temperature expansion of $Z_\text{Ising}$ in 
\eqref{eq:low-loop}. Since 
the closed loop in \eqref{eq:low-loop} corresponds to the worldline of $\psi$,
the low temperature expansion \eqref{eq:low-loop} can be recast as a fusion diagram.
As discussed in \cite{Aasen:2020jwb}, if we include the
small $\si$-loops around the vertices of the dual lattice, 
we can make this diagram trivalent.
A typical example of the fusion diagram looks like 
\footnote{See (3.2) in \cite{Aasen:2020jwb} as a comparison.}
\begin{equation}
\begin{aligned}
 \tikz[baseline={([yshift=-.5ex]current bounding box.center)}]{
\coordinate (a0) at (0,0); 
\coordinate (a1) at (1,0); 
\coordinate (a2) at (2,0); 
\coordinate (a3) at (3,0); 
\coordinate (s0) at (0.5,0.5);
\coordinate (s1) at (1.5,0.5); 
\coordinate (s2) at (2.5,0.5); 
\coordinate (s3) at (3.5,0.5); 
\coordinate (b0) at (0,1); 
\coordinate (b1) at (1,1); 
\coordinate (b2) at (2,1); 
\coordinate (b3) at (3,1); 
\coordinate (t0) at (0.5,1.5);
\coordinate (t1) at (1.5,1.5); 
\coordinate (t2) at (2.5,1.5); 
\coordinate (t3) at (3.5,1.5); 
\coordinate (c0) at (0,2); 
\coordinate (c1) at (1,2); 
\coordinate (c2) at (2,2); 
\coordinate (c3) at (3,2); 
\coordinate (u0) at (0.5,2.5);
\coordinate (u1) at (1.5,2.5); 
\coordinate (u2) at (2.5,2.5); 
\coordinate (u3) at (3.5,2.5);
\draw[dashed] (a0)--(s0)--(a1)--(s1)--(a2)--(s2)--(a3)--(s3);
\draw[dashed] (b0)--(s0)--(b1)--(s1)--(b2)--(s2)--(b3)--(s3);
\draw[dashed] (b0)--(t0)--(b1)--(t1)--(b2)--(t2)--(b3)--(t3);
\draw[dashed] (c0)--(t0)--(c1)--(t1)--(c2)--(t2)--(c3)--(t3);
\draw[dashed] (c0)--(u0)--(c1)--(u1)--(c2)--(u2)--(c3)--(u3);
\draw[green,thick] 
(s0) circle (0.2) (s1) circle (0.2) (s2) circle (0.2) (s3) circle (0.2)
(t0) circle (0.2) (t1) circle (0.2) (t2) circle (0.2) (t3) circle (0.2)
(u0) circle (0.2) (u1) circle (0.2) (u2) circle (0.2) (u3) circle (0.2); 
\fill 
(a0) circle (0.07) (a1) circle (0.07) (a2) circle (0.07) (a3) circle (0.07) 
(b0) circle (0.07) (b1) circle (0.07) (b2) circle (0.07) (b3) circle (0.07) 
(c0) circle (0.07) (c1) circle (0.07) (c2) circle (0.07) (c3) circle (0.07);
\draw[thick,red] (1.5,2.3)--(1.5,1.7);
\draw[thick,red] (1.7,1.5)--(2.3,1.5);
\draw[thick,red] (2.5,1.3)--(2.5,0.7);
\draw[thick,red] (2.3,0.5)--(1.7,0.5);
\draw[thick,red] (1.3,0.5)--(0.7,0.5);
\draw[thick,red] (0.5,2.3)--(0.5,1.7);
\draw[thick,red] (0.5,1.3)--(0.5,0.7);
\draw[thick,red] (1.3,2.5)--(0.7,2.5);
}
\end{aligned} 
\label{eq:fusion-lattice}
\end{equation}
where the red lines represent the domain wall labeled by $\psi$
and the green circles are the $\si$-loops.
If we are only interested in the low temperature expansion \eqref{eq:low-loop},
the green circles in \eqref{eq:fusion-lattice} are not necessary.
However, by adding them the diagram becomes trivalent,
which allows us to use the powerful technique of the fusion category.
We should stress that we are considering the trivalent diagram on a square lattice.
Sometimes, people consider a hexagonal lattice in order to draw a trivalent 
diagram on the lattice. 
However, it is not necessary;
in the AFM construction, we can draw a trivalent diagram
even on a square lattice as shown in \eqref{eq:fusion-lattice}.
\footnote{See also \cite{Hung:2025gcp} for the AFM construction
on a square lattice.}

As discussed in \cite{Aasen:2020jwb}, it is natural to 
assign the Boltzmann weight to a diamond
region bounded by the dashed lines in \eqref{eq:fusion-lattice}.
The Boltzmann weight are given by 
the product of two $6j$-symbols, each associated with 
the trivalent junction
\begin{equation}
\begin{aligned}
 \tikz[baseline={([yshift=-.5ex]current bounding box.center)},scale=1.6]{
\draw[thick,dashed] (0,0)--(0.5,0.5)--(1,0)--(0.5,-0.5)--(0,0);
\node at (0,0) [left] {$a$};
\node at (1,0) [right] {$b$};
\fill (0,0) circle (0.05) (1,0) circle (0.05);
\draw[thick,green] (0.25,0.25) to [out=-45, in=225] (0.75,0.25);
\draw[thick,green] (0.25,-0.25) to [out=45, in=135] (0.75,-0.25);
\draw[thick] (0.5,0.16) to (0.5,-0.16);
\node at (0.5,0) [right] {$\chi$};
}
=~
\tikz[baseline={([yshift=-.5ex]current bounding box.center)},scale=1.6]{
\draw[thick,dashed] (0,0)--(0.5,0.5)--(1,0)--(0.5,-0.5)--(0,0);
\node at (0.5,0.5) [above] {$a$};
\node at (0.5,-0.5) [below] {$b$};
\fill (0.5,0.5) circle (0.05) (0.5,-0.5) circle (0.05);
\draw[thick,green] (0.25,0.25) to [out=-45, in=45] (0.25,-0.25);
\draw[thick,green] (0.75,0.25) to [out=225, in=135] (0.75,-0.25);
\draw[thick] (0.34,0) to (0.66,0);
\node at (0.5,0) [above] {$\chi$};
}~
=\frac{1}{\rt{d_\si}}\bmat{\si~&\si~&\chi\\a~&b~&\si}
\bmat{\si~&\si~&\chi\\a~&b~&\si}
=\frac{\cob_{\chi,|a-b|}}{\rt{d_\si}}.
\end{aligned} 
\label{eq:diamond}
\end{equation}
Here, $a,b=0,1$ are the height variables related to the spin by \eqref{eq:sv-a},
while $\chi=0,1$ labels the fusion channel of 
$\si\otimes\si=\id+\psi$  
\begin{equation}
\begin{aligned}
 \chi=0~~&\Leftrightarrow~~\id,\\
\chi=1~~&\Leftrightarrow~~\psi.
\end{aligned} 
\end{equation}
The normalization $1/\rt{d_\si}$ in \eqref{eq:diamond}
can be removed by the ``bubble removal'' identity \eqref{eq:removal},
and hence this factor does not contribute to the final result.
For the nearest neighbor spins $s_v,s_{v'}$
\begin{equation}
\begin{aligned}
 s_v=(-1)^{a},\quad s_{v'}=(-1)^b,
\end{aligned} 
\end{equation}
the weight $e^{Js_vs_{v'}}$ is expanded as
\begin{equation}
\begin{aligned}
 e^{Js_vs_{v'}}=e^{J}\times\bigl(\cob_{|a-b|,0}+e^{-2J}\cob_{|a-b|,1}\bigr).
\end{aligned}
\label{eq:weight-cob} 
\end{equation}
One can see that this weight \eqref{eq:weight-cob} 
is written as a combination of the diamond weight in \eqref{eq:diamond}.
Thus, it is natural to define the Boltzmann weight as
\begin{equation}
W(v,v')=\left\{
\begin{aligned}
&\tikz[baseline={([yshift=-.5ex]current bounding box.center)},scale=1.6]{
\draw[thick,dashed] (0,0)--(0.5,0.5)--(1,0)--(0.5,-0.5)--(0,0);
\node at (0,0) [left] {$a$};
\node at (1,0) [right] {$b$};
\fill (0,0) circle (0.05) (1,0) circle (0.05);
\draw[thick,green] (0.25,0.25) to [out=-45, in=225] (0.75,0.25);
\draw[thick,green] (0.25,-0.25) to [out=45, in=135] (0.75,-0.25);
}~
+e^{-2J}~
\tikz[baseline={([yshift=-.5ex]current bounding box.center)},scale=1.6]{
\draw[thick,dashed] (0,0)--(0.5,0.5)--(1,0)--(0.5,-0.5)--(0,0);
\node at (0,0) [left] {$a$};
\node at (1,0) [right] {$b$};
\fill (0,0) circle (0.05) (1,0) circle (0.05);
\draw[thick,green] (0.25,0.25) to [out=-45, in=225] (0.75,0.25);
\draw[thick,green] (0.25,-0.25) to [out=45, in=135] (0.75,-0.25);
\draw[thick,red] (0.5,0.16) to (0.5,-0.16);
},
\qquad\bigl(\text{if}~\bra v,v'\ket~\text{is~horizontal}\bigr),\\
&\quad \tikz[baseline={([yshift=-.5ex]current bounding box.center)},scale=1.6]{
\draw[thick,dashed] (0,0)--(0.5,0.5)--(1,0)--(0.5,-0.5)--(0,0);
\node at (0.5,0.5) [above] {$a$};
\node at (0.5,-0.5) [below] {$b$};
\fill (0.5,0.5) circle (0.05) (0.5,-0.5) circle (0.05);
\draw[thick,green] (0.25,0.25) to [out=-45, in=45] (0.25,-0.25);
\draw[thick,green] (0.75,0.25) to [out=225, in=135] (0.75,-0.25);
}
~~~~+e^{-2J}~~~~
\tikz[baseline={([yshift=-.5ex]current bounding box.center)},scale=1.6]{
\draw[thick,dashed] (0,0)--(0.5,0.5)--(1,0)--(0.5,-0.5)--(0,0);
\node at (0.5,0.5) [above] {$a$};
\node at (0.5,-0.5) [below] {$b$};
\fill (0.5,0.5) circle (0.05) (0.5,-0.5) circle (0.05);
\draw[thick,green] (0.25,0.25) to [out=-45, in=45] (0.25,-0.25);
\draw[thick,green] (0.75,0.25) to [out=225, in=135] (0.75,-0.25);
\draw[thick,red] (0.34,0) to (0.66,0);
}\quad ,
\quad\quad\bigl(\text{if}~\bra v,v'\ket~\text{is~vertical}\bigr),
\end{aligned} \right.
\label{eq:Wvv'}
\end{equation}
where we have suppressed the overall normalization $e^J$ in 
\eqref{eq:weight-cob} for simplicity.
Note that the Boltzmann weights \eqref{eq:Wvv'}
for the horizontal and the vertical fusion channels are equal
since we assumed that the Ising coupling $J$ is homogeneous. 

Finally, the partition function $Z_\text{Ising}$ is written as
\begin{equation}
\begin{aligned}
 Z_\text{Ising}=\sum_{\{a_v=0,1\}}\prod_{\bra v,v'\ket}W(v,v').
\end{aligned} 
\label{eq:Z-height}
\end{equation}
As discussed in \cite{Aasen:2020jwb}, this partition function 
\eqref{eq:Z-height} has a holographic description in terms of the
Turaev-Viro state sum on a 3-manifold $T^2\times I$, where $I=[0,1]$ denotes the
interval. After summing over the height variables, 
\eqref{eq:Z-height} is written as a sum over the fusion diagrams $\cF$
\begin{equation}
\begin{aligned}
 Z_\text{Ising}=\sum_{\cF}A(\cF)\bra 0|\Pi|\cF\ket.
\end{aligned} 
\label{eq:Z-sumF}
\end{equation}
Here $\bra 0|\Pi|\cF\ket$ denotes the Turaev-Viro state sum, with 
$\Pi$ being some projector $\Pi^2=\Pi$. $A(\cF)$ in \eqref{eq:Z-sumF}
is an amplitude of the fusion diagram $\cF$ \cite{Aasen:2020jwb};
in the case of the Ising model, it is given by
\begin{equation}
\begin{aligned}
 A(\cF)=(e^{-2J})^{N_\psi(\cF)},
\end{aligned} 
\end{equation}
where $N_\psi(\cF)$ is the number of $\psi$-lines in the diagram $\cF$.

More generally, 
we can introduce the macroscopic defects $\cD$ into the model.
Then the partition function in the presence of defects $\cD$ becomes
\begin{equation}
\begin{aligned}
 Z_\text{Ising}(\cD)=\sum_{\cF}A(\cF)\bra \cD|\Pi|\cF\ket.
\end{aligned} 
\label{eq:Z-D}
\end{equation}
The overlap $\bra\cD|\Pi|\cF\ket$ has a natural interpretation 
as a sandwich construction of SymTFT.
In section \ref{sec:holography}, we will elaborate more details on the holographic 
interpretation of the partition function \eqref{eq:Z-D} with defects $\cD$.

As discussed in \cite{Aasen:2016dop,Aasen:2020jwb}, the defect operators $\cD_a$
are labeled by a simple object $a\in\text{Irr}(\cC)$ and they obey the same fusion
algebra as \eqref{eq:fusion-N}
\begin{equation}
\begin{aligned}
 \cD_a\cD_b=\sum_{c\in\text{Irr}(\cC)}N_{ab}^c\cD_c.
\end{aligned} 
\end{equation}
For the Ising case, $\cD_\si$ represents the 
Kramers-Wannier duality defect, while $\cD_\psi$ is the spin-flip defect 
where the spins on the two sides of the defect $\cD_\psi$
have opposite signs.
From \eqref{eq:fusion-Ising}, $\cD_\si$ and $\cD_\psi$ obey the following
fusion rule
\begin{equation}
\begin{aligned}
 \cD_\psi^2=\cD_\id,\quad \cD_\si^2=\cD_\id+\cD_\psi,\quad
\cD_\psi\cD_\si=\cD_\si.
\end{aligned} 
\label{eq:D-fusion}
\end{equation}

\section{Partition function of Ising CFT}\label{sec:CFT}
In the continuum limit at the critical point $J=J_c$, the Ising model
is described by the $c=1/2$ CFT.
The partition function $Z_\text{Ising}$ of the Ising CFT
is given by the diagonal combination of the characters
\begin{equation}
\begin{aligned}
 Z_\text{Ising}=|\chi_\id|^2+|\chi_\psi|^2+|\chi_\si|^2,
\end{aligned} 
\label{eq:Z-Isng-CFT}
\end{equation}
where the three characters are given by
\begin{equation}
\begin{aligned}
 \chi_\id(\tau)&=\hf q^{-\frac{1}{48}}\left[\prod_{n=1}^\infty (1+q^{n-\hf})
+\prod_{n=1}^\infty (1-q^{n-\hf})\right],\\
\chi_\psi(\tau)&=\hf q^{-\frac{1}{48}}\left[\prod_{n=1}^\infty (1+q^{n-\hf})
-\prod_{n=1}^\infty (1-q^{n-\hf})\right],\\
\chi_\si(\tau)&=q^{\frac{1}{24}}\prod_{n=1}^\infty (1+q^n),
\end{aligned} 
\end{equation}
with $q=e^{2\pi\ri\tau}$.
Introducing the vector $v(\tau)$
\begin{equation}
\begin{aligned}
 v(\tau)=\mat{\chi_\id(\tau)\\\chi_\si(\tau)\\\chi_\psi(\tau)},
\end{aligned} 
\label{eq:v-tau}
\end{equation}
the $S$- and $T$-transformation are written as
\begin{equation}
\begin{aligned}
 v(-1/\tau)=Sv(\tau),\quad v(\tau+1)=e^{-\frac{\pi\ri}{24}}Tv(\tau), 
\end{aligned} 
\end{equation}
where the $S$-matrix and $T$-matrix are given by
\begin{equation}
\begin{aligned}
 S=\hf\mat{1&\rt{2}&1\\
\rt{2}&0&-\rt{2}\\
1&-\rt{2}&1
},\quad
T=\mat{~1~ & &\\ 
&~\th_\si~& \\ && ~-1~},
\end{aligned} 
\label{eq:S-T}
\end{equation}
with $\th_\si$ being
\begin{equation}
\begin{aligned}
 \th_\si=e^{\frac{2\pi\ri}{16}}.
\end{aligned} 
\label{eq:th-si}
\end{equation}

In general, the partition function with defects are written as a linear 
combination
of $v(\tau)$ and its complex conjugate.
A convenient basis of the partition functions is 
\begin{equation}
\begin{aligned}
 \cZ_M(\tau)=v(\tau)^\dag M v(\tau),
\end{aligned} 
\label{eq:ZM}
\end{equation}
where $M$ is a $3\times 3$ matrix.
As discussed in \cite{Aasen:2016dop},
there are 10 independent partition functions with defects
\begin{equation}
\begin{aligned}
 \Zemp&=\b{\chi}_\id\chi_\id+\b{\chi}_\psi\chi_\psi+\b{\chi}_\si\chi_\si,\\
\Zhp&=\b{\chi}_\id\chi_\id+\b{\chi}_\psi\chi_\psi-\b{\chi}_\si\chi_\si,\\
\Zhs&=\rt{2}\b{\chi}_\id\chi_\id-\rt{2}\b{\chi}_\psi\chi_\psi,\\
\Zvp&=\b{\chi}_\id\chi_\psi+\b{\chi}_\psi\chi_\id+\b{\chi}_\si\chi_\si,\\
\Zvs&=(\b{\chi}_\id+\b{\chi}_\psi)\chi_\si+\b{\chi}_\si(\chi_\id+\chi_\psi),\\
\Ztp&=\b{\chi}_\si\chi_\si-\b{\chi}_\id\chi_\psi-\b{\chi}_\psi\chi_\id,\\
\Zts&=(\b{\chi}_\id-\b{\chi}_\psi)\th_\si\chi_\si+\b{\chi}_\si\b{\th}_\si
(\chi_\id-\chi_\psi),\\
\Ztsr&=(\b{\chi}_\id-\b{\chi}_\psi)\b{\th}^3_\si\chi_\si+\b{\chi}_\si\th_\si^3
(\chi_\id-\chi_\psi),\\
\Zsp&=(\b{\chi}_\id+\b{\chi}_\psi)\ri\chi_\si-\ri\b{\chi}_\si (\chi_\id+\chi_\psi),
\\
\Zps&=\rt{2}\ri\bigl(\b{\chi}_\id\chi_\psi-\b{\chi}_\psi\chi_\id\bigr).
\end{aligned} 
\label{eq:Z-defect-10}
\end{equation}
Above, we have suppressed the argument $\tau$ for brevity.
The square in \eqref{eq:Z-defect-10} schematically
depicts the torus, where the left-right and the top-bottom sides of the square are identified.
The green lines and the red lines represent the defects
$\cD_\si$ and $\cD_\psi$, respectively. 
Note that the combinations in \eqref{eq:Z-defect-10} are all real.
Among \eqref{eq:Z-defect-10}, there is one non-trivial relation
\begin{equation}
\begin{aligned}
 \Zemp=\hf\left(\Zemp+\Zhp+\Zvp+\Ztp\right),
\end{aligned} 
\end{equation}
which comes from the fact 
that the Ising model is equivalent to its
$\mathbb{Z}_2$ gauging at criticality $J=J_c$
\cite{Shao:2023gho,Kaidi:2026urc} \footnote{More generally,
the Ising model with coupling $\til{J}$ is related to
its $\mathbb{Z}_2$-gauging at coupling $J$
\cite{Kaidi:2026urc}
\begin{equation}
\begin{aligned}
 \text{Ising}\,[\til{J}]=\text{Ising}/\mathbb{Z}_2\,[J].
\end{aligned} 
\end{equation}
}
\begin{equation}
\begin{aligned}
 \text{Ising}~=~\text{Ising}/\mathbb{Z}_2~.
\end{aligned} 
\end{equation}
Inverting the relations \eqref{eq:Z-defect-10}, we find
that the product of characters $\b{\chi}_c\chi_d$ are written as
\begin{equation}
\begin{aligned}
\b{\chi}_\id\chi_\id&=
\qu\left(\Zemp+\Zhp+\rt{2}~\Zhs\right),\\
\b{\chi}_\psi\chi_\psi&=
\qu\left(\Zemp+\Zhp-\rt{2}~\Zhs\right),\\
\b{\chi}_\si\chi_\si&=\hf\left(\Zemp-\Zhp\right)=\hf\left(\Zvp+\Ztp\right),\\
 \b{\chi}_\id\chi_\psi&=\qu\left(~\Zvp-\Ztp-\rt{2}\ri~\Zps~\right),\\
\b{\chi}_\psi\chi_\id&=
\qu\left(~\Zvp-\Ztp+\rt{2}\ri~\Zps~\right),\\
\b{\chi}_\id\chi_\si&=\qu\left(~\Zvs-\ri~\Zsp+
\th_\si^3~\Ztsr+\b{\th}_\si~\Zts~\right),\\
\b{\chi}_\si\chi_\id&=
\qu\left(~\Zvs+\ri~\Zsp+
\b{\th}_\si^3~\Ztsr+\th_\si~\Zts~\right),\\
\b{\chi}_\psi\chi_\si&=\qu\left(~\Zvs-\ri~\Zsp-\th_\si^3~\Ztsr-\b{\th}_\si~\Zts~\right),\\
\b{\chi}_\si\chi_\psi&=
\qu\left(~\Zvs+\ri~\Zsp-\b{\th}_\si^3~\Ztsr-\th_\si~\Zts~\right).
\end{aligned} 
\label{eq:chi-chi-Z}
\end{equation}
As we will see in section \ref{sec:tube},
the above expressions \eqref{eq:chi-chi-Z} are closely related to the 
central idempotents in the tube algebra $\text{Tube}(\cC)$.

For instance, let us consider the idempotent associated with
the diagonal character $\b{\chi}_a\chi_a$ in \eqref{eq:chi-chi-Z}. 
To this end, it is convenient to introduce the total Hilbert space
$\cH_\text{total}$ on a constant Euclidean time slice of the torus,
i.e. the Hilbert space on a circle $S^1$.
As discussed in \cite{Lin:2022dhv}, $\cH_\text{total}$
is decomposed into the twisted sectors $\cH_a$ labeled by a simple object
$a\in\text{Irr}(\cC)$
\begin{equation}
\begin{aligned}
 \cH_\text{total}=\bigoplus_{a\in\text{Irr}(\cC)}\cH_a,
\end{aligned} 
\label{eq:H-tot}
\end{equation}
and $\cH_a$ is further decomposed as a product of the left-moving $\b{V}_c$
and the right-moving $V_d$ irreducible representations
of the Virasoro algebra
\begin{equation}
\begin{aligned}
 \cH_a=\bigoplus_{c,d}N_{ac}^d\b{V}_c\otimes V_d.
\end{aligned} 
\end{equation}
For the Ising model, there are three twisted sectors $\cH_\id,\cH_\psi,\cH_\si$
\begin{equation}
\begin{aligned}
 \cH_\id&=\b{V}_\id\otimes V_\id+\b{V}_\psi\otimes V_\psi
+\b{V}_\si\otimes V_\si,\\
\cH_\psi&=\b{V}_\id\otimes V_\psi+\b{V}_\psi\otimes V_\id+ \b{V}_\si\otimes V_\si,\\
\cH_\si&=(\b{V}_\id+\b{V}_\psi)\otimes V_\si+\b{V}_\si\otimes (V_\id+V_\psi).
\end{aligned} 
\label{eq:Ha-Ising}
\end{equation}
It turns out that the diagonal character $\b{\chi}_a\chi_a$
is written as a trace on $\cH_\id$
with the insertion of the projector $P_\id^{(a,a)}$
\begin{equation}
\begin{aligned}
 \b{\chi}_a\chi_a=\Tr_{\cH_\id}\Bigl(P_\id^{(a,a)}
\b{q}^{\b{L}_0-\frac{c}{24}}q^{L_0-\frac{c}{24}}\Bigr).
\end{aligned} 
\end{equation}
As shown in \cite{Lin:2022dhv}, the projector $P_\id^{(a,a)}$
is written as a combination of the defects $\cD_b$
\begin{equation}
\begin{aligned}
 P_\id^{(a,a)}=S_{\id a}\sum_{b\in\text{Irr}(\cC)} \b{S_{ba}}\cD_b,
\end{aligned} 
\label{eq:P-id}
\end{equation}
where the defect line $\cD_b$ is running
along the horizontal direction on the torus.
For instance, the so-called Kirby element
$P^{(\id,\id)}_\id$ is written as
\begin{equation}
\begin{aligned}
 P^{(\id,\id)}_\id=\frac{1}{\text{dim}(\cC)}\sum_{b\in\text{Irr}(\cC)} d_b\cD_b,
\end{aligned} 
\end{equation}
where we used the relation
\begin{equation}
\begin{aligned}
 d_a=\frac{S_{\id a}}{S_{\id\id}},\qquad \frac{1}{S_{\id\id}}=\text{dim}(\cC)
=\sum_{a\in\text{Irr}(\cC)}d_a^2.
\end{aligned} 
\end{equation}
For the Ising case, the projectors $P_\id^{(a,a)}$ in \eqref{eq:P-id}
are given by
\begin{equation}
\begin{aligned}
 P_\id^{(\id,\id)}&=\frac{1}{4}\left(\cD_\id+\cD_\psi+\rt{2}\cD_\si\right),\\
P_\id^{(\psi,\psi)}&=\frac{1}{4}\left(\cD_\id+\cD_\psi-\rt{2}\cD_\si\right),\\
P_\id^{(\si,\si)}&=\frac{1}{2}\left(\cD_\id-\cD_\psi\right).
\end{aligned} 
\label{eq:P1-ising}
\end{equation}
One can easily see that the diagonal characters $\b{\chi}_a\chi_a$
in \eqref{eq:chi-chi-Z} are obtained by the insertion of the projectors
\eqref{eq:P1-ising}.
Using the fusion rule \eqref{eq:D-fusion},
one can also check that $P_\id^{(a,a)}$ in \eqref{eq:P1-ising}
are orthogonal projectors
\begin{equation}
\begin{aligned}
 P_\id^{(a,a)}P_\id^{(b,b)}=\cob^{a,b}P_\id^{(a,a)}.
\end{aligned} 
\end{equation}
More generally, we can define the projector
$P_a^{(c,d)}$ which projects the states of
$\cH_\text{total}$ onto the $\b{V}_c\otimes V_d$ part
of the twisted sector $\cH_a$
\begin{equation}
\begin{aligned}
 P_a^{(c,d)}\cH_\text{total}=N_{ac}^d \b{V}_c\otimes V_d~\subset~ \cH_a.
\end{aligned} 
\label{eq:P-acd}
\end{equation}
In section \ref{sec:tube},
we will consider a general formula of $P_a^{(c,d)}$
in terms of the tube algebra $\text{Tube}(\cC)$.

\section{Tube algebra and idempotents}\label{sec:tube}
In this section, we consider the tube algebra 
$\text{Tube}(\cC)$ \cite{ocneanu1994,evans1995ocneanu} and the idempotents (or projectors)
$P_a^{(c,d)}$ in \eqref{eq:P-acd}.
As discussed in \cite{Lin:2022dhv,kawagoe2024levin},
the tube algebra naturally arises if we consider the action of defect operators
on a local operator in the picture of the radial quantization 
and the state-operator correspondence.
In the radial quantization, a local operator
$\cO$ inserted at 
the center of a disk $D^2$ is mapped to a state $|\cO\ket$ in 
the Hilbert space $\cH_\text{total}$ 
\eqref{eq:H-tot} on the boundary circle $S^1=\del D^2$ of the disk. 
By the decomposition \eqref{eq:H-tot} into the twisted sector, the state
$|\cO\ket\in\cH_\text{total}$ is expanded as
\begin{equation}
\begin{aligned}
 |\cO\ket=\sum_{a\in\text{Irr}(\cC)}|\cO\ket_a,\qquad |\cO\ket_a\in\cH_a.
\end{aligned} 
\end{equation}
Then the action of the defect operator $\cD_c$ surrounding $\cO$
is represented by a contribution of the annular region (or tube)
containing $\cD_c$, and this tube naturally acts 
on the state $|\cO\ket_a$ in the
Hilbert space $\cH_a$ on the boundary circle $S^1=\del D^2$
as
\begin{equation}
\begin{aligned}
 \tikz[baseline={([yshift=-.5ex]current bounding box.center)},scale=1.6]{
\draw[fill=magenta,opacity=0.2] (0,0) circle (0.3);
\draw[thick] (0,0) circle (0.3);
\node at (0,0) {$\cO$};
\draw[thick,blue] (0,0) circle (0.7);
\node at (0.67,0) [right] {$c$};
\draw[thick] (0,0) circle (1.1);
}
~=~\bigoplus_{a,b\in\text{Irr}(\cC)}~
\tikz[baseline={([yshift=-.5ex]current bounding box.center)}]{
\draw[thick] (1,2) arc (0:360: 1 and 0.4); 
\draw[fill=magenta,opacity=0.2] (1,-1) arc (0:360: 1 and 0.4); 
\draw[thick] (-1,-1) arc (180:360: 1 and 0.4); 
\draw[thick] (1,-1) -- (1,2);
\draw[thick] (-1,-1) -- (-1,2);
\node at (1,-1) [right] {$\cH_a$};
\node at (1,2) [right] {$\cH_b$};
\draw[thick,blue] (-1,0.5) arc (180:360: 1 and 0.4); 
\draw[thick,dotted,blue] (1,0.5) arc (0:180: 1 and 0.4);
\draw[fill,white] (0,0.1) circle (0.3);
\draw[thick,blue] (0,0.1) circle (0.3);
\node at (0,0.1) {$f$};
\draw[thick,blue] (0,-1.4) -- (0,-0.2);
\draw[thick,blue] (0,0.4) -- (0,1.6);
\node at (0,-1.4) [below] {$a$};
\node at (0,1.6) [above] {$b$};
\node at (0.6,0.13) [above] {$c$};
\node at (-0.6,0.13) [above] {$c$};
}.
\end{aligned} 
\label{eq:tube-fig}
\end{equation} 
By reading the figure on the right-hand side form SE (southeast) to NW 
(northwest), the symbol $f$ in \eqref{eq:tube-fig} is understood as
\begin{equation}
\begin{aligned}
 f\in\text{Hom}_\cC(c\otimes a\to b\otimes c).
\end{aligned} 
\end{equation}
This $f$ is an element of the tube algebra
defined by
\begin{equation}
\begin{aligned}
 \text{Tube}(\cC)=\bigoplus_{a,b,c\in\text{Irr}(\cC)}
\text{Hom}_\cC(c\otimes a\to b\otimes c).
\end{aligned} 
\end{equation}
The multiplication in the tube algebra is naturally realized
by the stacking of tubes \cite{kawagoe2024levin}.

As discussed in \cite{Lin:2022dhv,Williamson:2017uzx,Bultinck:2015bot},
we can construct the idempotent $P_a^{(c,d)}$ in \eqref{eq:P-acd}
as an element of $\text{Tube}(\cC)$.
Note that only the $a=b$ case in \eqref{eq:tube-fig}
is relevant in the construction of the idempotent $P_a^{(c,d)}$ in \eqref{eq:P-acd}.
Based on this observation, 
we introduce the basis of the tube algebra
\begin{equation}
\begin{aligned}
 \cT^c_{ab}=
\tikz[baseline={([yshift=-.5ex]current bounding box.center)},scale=1.6]{
\draw[thick] (0,0)--(0,1);
\draw[thick] (1,0)--(1,1);
\draw[thick,dashed] (0,0)--(1,0);
\draw[thick,dashed] (0,1)--(1,1);
\draw[thick,blue] (0,0.5) to [out=0, in=-90] (0.5,1);
\draw[thick,blue] (0.5,0) to [out=90, in=180] (1,0.5);
\draw[thick,blue] (0.35,0.65)--(0.65,0.35);
\node at (0.5,0) [below] {$a$};
\node at (0.5,1) [above] {$a$};
\node at (1,0.5) [right] {$b$};
\node at (0,0.5) [left] {$b$};
\node at (0.6,0.4) [above] {$c$};
}~\in~\text{Hom}_\cC(b\otimes a\to a\otimes b),
\end{aligned} 
\end{equation}
where the left side and the right side of the square are identified.
The three-dimensional picture of $\cT^c_{ab}$ looks like
\begin{equation}
\begin{aligned}
 \cT^c_{ab}~=~
\tikz[baseline={([yshift=-.5ex]current bounding box.center)},scale=1.6]{
\draw[thick] (0.5,1) arc (0:360: 0.5 and 0.2); 
\draw[thick] (-0.5,0) arc (180:360: 0.5 and 0.2); 
\draw[thick,dotted] (0.5,0) arc (0:180: 0.5 and 0.2); 
\draw[thick] (0.5,0) -- (0.5,1);
\draw[thick] (-0.5,0) -- (-0.5,1);
\draw[thick,blue] (0,-0.2) to [out=90, in=250] (0.5,0.4);
\draw[thick,blue] (-0.5,0.4) to [out=-70, in=270] (0,0.8);
\draw[thick,dashed,blue] (0.5,0.4) arc (0:180: 0.5 and 0.2); 
\draw[thick,blue] (0.2,0.1) -- (-0.2,0.45);
\node at (0.08,0.2) [above] {$c$};
\node at (0,-0.2) [below] {$a$};
\node at (0,0.75) [above] {$a$};
\node at (0.5,0.4) [right] {$b$};
\node at (-0.5,0.4) [left] {$b$};
}.
\end{aligned} 
\end{equation}

We will give a general formula of the idempotent
in terms of the braid $R$-symbol $R_c^{ab}$ defined by
\begin{equation}
\begin{aligned}
\tikz[baseline={([yshift=-.5ex]current bounding box.center)},scale=1.6]{
\draw[thick] (-0.2,0.2) arc (180:360:0.2);
\draw[thick] (0,0) -- (0,-0.3);
\draw[thick] (0.2,0.2) to [out=90,in=-90] (-0.2,0.8);
\draw[line width=5pt, white] (-0.2,0.2) to [out=90,in=-90] (0.2,0.8);
\draw[thick] (-0.2,0.2) to [out=90,in=-90] (0.2,0.8);
\node at (0,-0.3) [below] {$c$};
\node at (-0.2,0.8) [above] {$b$};
\node at (0.2,0.8) [above] {$a$};
}
~=~
R^{ab}_c
\tikz[baseline={([yshift=-.5ex]current bounding box.center)},scale=1.6]{
\draw[thick] (-0.2,0.2) arc (180:360:0.2);
\draw[thick] (0,0) -- (0,-0.3);
\draw[thick] (-0.2,0.2) -- (-0.2,0.7);
\draw[thick] (0.2,0.2) -- (0.2,0.7);
\node at (0,-0.3) [below] {$c$};
\node at (-0.2,0.7) [above] {$b$};
\node at (0.2,0.7) [above] {$a$};
}~,\qquad
\tikz[baseline={([yshift=-.5ex]current bounding box.center)},scale=1.6]{
\draw[thick] (-0.2,0.2) arc (180:360:0.2);
\draw[thick] (0,0) -- (0,-0.3);
\draw[thick] (-0.2,0.2) to [out=90,in=-90] (0.2,0.8);
\draw[line width=5pt,white] (0.2,0.2) to [out=90,in=-90] (-0.2,0.8);
\draw[thick] (0.2,0.2) to [out=90,in=-90] (-0.2,0.8);
\node at (0,-0.3) [below] {$c$};
\node at (-0.2,0.8) [above] {$b$};
\node at (0.2,0.8) [above] {$a$};
}
~=~
\b{R}^{ab}_c
\tikz[baseline={([yshift=-.5ex]current bounding box.center)},scale=1.6]{
\draw[thick] (-0.2,0.2) arc (180:360:0.2);
\draw[thick] (0,0) -- (0,-0.3);
\draw[thick] (-0.2,0.2) -- (-0.2,0.7);
\draw[thick] (0.2,0.2) -- (0.2,0.7);
\node at (0,-0.3) [below] {$c$};
\node at (-0.2,0.7) [above] {$b$};
\node at (0.2,0.7) [above] {$a$};
}~.
\end{aligned}
\label{eq:R-move}
\end{equation}
For the Ising model, the $R$-symbols are given by 
(see Table 1 in \cite{Kitaev:2005hzj})
\begin{equation}
\begin{aligned}
 &R^{\psi\psi}_\id=-1,\quad R^{\si\si}_\id=\b{\th}_\si,
\quad R^{\si\psi}_\si=R^{\psi\si}_\si=-\ri,\quad
R^{\si\si}_\psi=\th_\si^3,\\
&R^{ab}_c=N_{ab}^c~~(\text{otherwise}).
\end{aligned} 
\label{eq:R-Ising}
\end{equation}

Now, let us consider the projectors
$P_a^{(c,*)}$ and $P_a^{(*,d)}$
\begin{equation}
\begin{aligned}
 P_a^{(c,*)}\cH_\text{total}&=\bigoplus_{d\in\text{Irr}(\cC)}N_{ac}^d \b{V}_c\otimes V_d,\\
 P_a^{(*,d)}\cH_\text{total}&=
\bigoplus_{c\in\text{Irr}(\cC)}N_{ac}^d \b{V}_c\otimes V_d.
\end{aligned} 
\end{equation}
As discussed in \cite{Lin:2022dhv},
the projector $P_a^{(c,*)}$
is given by the crossed diagram of the defect lines $\cD_a$ and $\cD_b$
(also known as the Verlinde lines \cite{Verlinde:1988sn}) 
\begin{equation}
\begin{aligned}
 P_a^{(c,*)}=\sum_{b\in\text{Irr}(\cC)}S_{\id c}\b{S}_{bc}~
\tikz[baseline={([yshift=-.5ex]current bounding box.center)}]{
\draw[thick,blue] (-1,0.7)--(1,0.7);
\draw[line width=4pt,white] (0,0)--(0,1.5);
\draw[thick,blue] (0,0)--(0,1.5);
\node at (1,0.7) [right] {$b$};
\node at (0,0) [below] {$a$};
\node at (0,1.5) [above] {$a$};
},
\end{aligned} 
\label{eq:P-cross}
\end{equation}
where the left side and the right side of the figure are understood to be identified.
Using \eqref{eq:cup-cap} and \eqref{eq:R-move}, we can rewrite
\eqref{eq:P-cross} as
\begin{equation}
\begin{aligned}
  P_a^{(c,*)}&=\sum_{b,c'\in\text{Irr}(\cC)}S_{\id c}\b{S}_{bc}
\rt{\frac{d_{c'}}{d_ad_b}}~
\tikz[baseline={([yshift=-.5ex]current bounding box.center)},scale=1.6]{
\draw[thick,blue] (0.5,0) to [out=90,in=180] (1,0.5);
\draw[thick,blue] (0.64,0.33) -- (0.3,0.65);
\draw[thick,blue] (0.3,0.65) arc (-45:90:0.3);
\draw[thick,blue] (0.3,0.65) arc (-45:-180:0.3);
\draw[thick,blue] (-0.7,1.165) -- (0.1,1.165);
\draw[line width=4pt,white] (-0.213,0.84) -- (-0.213,1.5);
\draw[thick,blue] (-0.213,0.84) -- (-0.213,1.7);
\node at (0.5,0) [below] {$a$};
\node at (1,0.5) [right] {$b$};
\node at (-0.213,1.7) [above] {$a$};
\node at (-0.7,1.165) [left] {$b$};
\node at (0.6,0.4) [above] {$c'$};
}\\
&=\sum_{b,c'\in\text{Irr}(\cC)}S_{\id c}\b{S}_{bc}
\rt{\frac{d_{c'}}{d_ad_b}}R^{ab}_{c'}~
\tikz[baseline={([yshift=-.5ex]current bounding box.center)},scale=1.6]{
\draw[thick,blue] (0,0.5) to [out=0, in=-90] (0.5,1);
\draw[thick,blue] (0.5,0) to [out=90, in=180] (1,0.5);
\draw[thick,blue] (0.35,0.65)--(0.65,0.35);
\node at (0.5,0) [below] {$a$};
\node at (0.5,1) [above] {$a$};
\node at (1,0.5) [right] {$b$};
\node at (0,0.5) [left] {$b$};
\node at (0.6,0.45) [above] {$c'$};
}\\
&=\sum_{b,c'\in\text{Irr}(\cC)}S_{\id c}\b{S}_{bc}
\rt{\frac{d_{c'}}{d_ad_b}}R^{ab}_{c'}\cT^{c'}_{ab}.
\end{aligned} 
\label{eq:P-R}
\end{equation}
As far as we know, this expression \eqref{eq:P-R}
of $P_a^{(c,*)}$ in terms of the $R$-symbol is not known 
in the literature before.
Similarly, $P_a^{(*,d)}$ is written as
\begin{equation}
\begin{aligned}
P_a^{(*,d)}&= 
\sum_{b\in\text{Irr}(\cC)}S_{\id d}\b{S}_{bd}~
\tikz[baseline={([yshift=-.5ex]current bounding box.center)}]{
\draw[thick,blue] (0,0)--(0,1.5);
\draw[line width=4pt,white] (-1,0.7)--(1,0.7);
\draw[thick,blue] (-1,0.7)--(1,0.7);
\node at (1,0.7) [right] {$b$};
\node at (0,0) [below] {$a$};
\node at (0,1.5) [above] {$a$};
}
=\sum_{b,c\in\text{Irr}(\cC)}S_{\id d}\b{S}_{bd}
\rt{\frac{d_{c}}{d_ad_b}}\b{R}^{ab}_{c}\cT^{c}_{ab}.
\end{aligned} 
\label{eq:P-barR}
\end{equation}
We stress that our formula \eqref{eq:P-R} and \eqref{eq:P-barR}
can be applied to arbitrary modular tensor category $\cC$.
Finally, the minimal central idempotent $P^{(c,d)}$ is given by
\begin{equation}
\begin{aligned}
 P^{(c,d)}=\sum_{a\in\text{Irr}(\cC)}P_a^{(c,*)}P_a^{(*,d)},
\end{aligned} 
\label{eq:P-min}
\end{equation}
which projects to the factor $\b{V}_c\otimes V_d$ in $\cH_\text{total}$
\begin{equation}
\begin{aligned}
 P^{(c,d)}\cH_\text{total}=\bigoplus_{a\in\text{Irr}(\cC)}N_{ac}^d\b{V}_c\otimes V_d.
\end{aligned} 
\end{equation}

For the Ising model,
we can explicitly write down the projectors
using the $R$-symbol \eqref{eq:R-Ising}
and the $S$-matrix \eqref{eq:S-T} of the Ising model.
Because of the structure of the Hilbert space 
\eqref{eq:Ha-Ising}, we do not need to multiply
$P_a^{(c,*)}$ and $P_a^{(*,d)}$ to extract the factor 
$\b{V}_c\otimes V_d$.
For instance, we find the relations
\begin{equation}
\begin{aligned}
 P_\psi^{(\id,*)}\cH_\psi&=\b{V}_\id\otimes V_\psi,\quad
&P_\psi^{(*,\id)}\cH_\psi&=\b{V}_\psi\otimes V_\id\\
P_\si^{(\id,*)}\cH_\si&=\b{V}_\id\otimes V_\si,\quad
&P_\si^{(*,\id)}\cH_\si&=\b{V}_\si\otimes V_\id,\\
P_\si^{(\psi,*)}\cH_\si&=\b{V}_\psi\otimes V_\si,\quad
&P_\si^{(*,\psi)}\cH_\si&=\b{V}_\si\otimes V_\psi.
\end{aligned} 
\label{eq:Ising-simple-P}
\end{equation}
Applying the general formula \eqref{eq:P-R} and \eqref{eq:P-barR}
to the Ising model with the relations \eqref{eq:Ising-simple-P},
we find the minimal central idempotents
\begin{equation}
\begin{aligned}
 P^{(\id,\id)}&=\qu\bigl(\cT^{\id}_{\id\id}+\cT^{\id}_{\id\psi}
+\rt{2}\cT^{\id}_{\id\si}\bigr),\\
P^{(\psi,\psi)}&=\qu\bigl(\cT^{\id}_{\id\id}
+\cT^{\id}_{\id\psi}
-\rt{2}\cT^{\id}_{\id\si}\bigr),\\
P^{(\si,\si)}&=\hf\bigl(\cT^{\id}_{\id\id}-\cT^{\id}_{\id\psi}
+\cT^{\id}_{\psi\id}+\cT^{\id}_{\psi\psi}\bigr),\\
P^{(\id,\psi)}&=\qu\bigl(\cT^{\id}_{\psi\id}-\cT^{\id}_{\psi\psi}
-\rt{2}\ri\cT^{\si}_{\psi\si}\bigr),\\
P^{(\psi,\id)}&=
\qu\bigl(\cT^{\id}_{\psi\id}-\cT^{\id}_{\psi\psi}
+\rt{2}\ri\cT^{\si}_{\psi\si}\bigr),\\
P^{(\id,\si)}&=\qu\bigl(\cT^{\id}_{\si\id}-\ri\cT^{\si}_{\si\psi}
+\th^3_\si\cT^{\psi}_{\si\si}+\b{\th}_\si\cT^{\id}_{\si\si}\bigr),\\
P^{(\si,\id)}&=\qu\bigl(\cT^{\id}_{\si\id}+\ri\cT^{\si}_{\si\psi}
+\b{\th}^3_\si\cT^{\psi}_{\si\si}+\th_\si\cT^{\id}_{\si\si}\bigr),\\
P^{(\psi,\si)}&=\qu\bigl(\cT^{\id}_{\si\id}-\ri\cT^{\si}_{\si\psi}
-\th^3_\si\cT^{\psi}_{\si\si}-\b{\th}_\si\cT^{\id}_{\si\si}\bigr),\\
P^{(\si,\psi)}&=\qu\bigl(\cT^{\id}_{\si\id}+\ri\cT^{\si}_{\si\psi}
-\b{\th}^3_\si\cT^{\psi}_{\si\si}-\th_\si\cT^{\id}_{\si\si}\bigr).
\end{aligned} 
\label{eq:P-Ising}
\end{equation}
This agrees with the result in \cite{Williamson:2017uzx,Bultinck:2015bot}, which was
obtained by the brute force method.
Our general formulas \eqref{eq:P-R} and \eqref{eq:P-barR}
give a natural interpretation for the coefficients 
in \eqref{eq:P-Ising} as the $R$-symbol of the Ising model \eqref{eq:R-Ising}.
The torus partition functions with defects in \eqref{eq:chi-chi-Z}
are reproduced by taking the trace of \eqref{eq:P-Ising}, since the torus
is obtained by identifying the top and bottom sides of the tube.

\section{Holographic dual of 2d Ising model}\label{sec:holography}
In this section, we consider a holographic dual of the 2d Ising model
on a torus.
We set the coupling to its critical value $J=J_c$,
but we will not take a continuum limit, i.e. we keep the 
lattice spacing finite.
We stress that our holographic duality works
for the Ising model on a lattice without taking the continuum limit.
 
We propose that the bulk dual of the 2d Ising model 
on a torus is a 3d Turaev-Viro TQFT on a solid torus
$D^2\times S^1$.
This solid torus is further decomposed into two pieces via the cobordism
\begin{equation}
\begin{aligned}
 \emptyset~\xrightarrow{~\text{solid~torus}~}
~T^2~\xrightarrow{~T^2\times I~}~T^2,
\end{aligned} 
\label{eq:cobordism}
\end{equation}
where $I=[0,1]$ denotes the interval.
We will call the first part of \eqref{eq:cobordism} the ``core'' 
and the second part of \eqref{eq:cobordism} the ``sandwich''.
For the sandwich region,
we call the two ends of the cylinder $T^2\times I$
the ``topological boundary'' and the
``physical boundary''
\begin{equation}
\begin{aligned}
 \text{topological boundary:}\quad\mathbb{D}&=T^2\times\{0\},\\
\text{physical boundary:}\quad\mathbb{B}&=T^2\times\{1\}.
\end{aligned} 
\label{eq:def-DB}
\end{equation}
Then \eqref{eq:cobordism} is written as
\begin{equation}
\begin{aligned}
 \text{core}:&\quad \emptyset~\xrightarrow{~\text{solid~torus}~}
~\mathbb{D},\\
\text{sandwich}:&\quad \mathbb{D}~\xrightarrow{~T^2\times I~}~\mathbb{B}.
\end{aligned} 
\label{eq:core-sandwich}
\end{equation}
We also introduce
the ``bedding surface'' $\mathbb{S}$ in the middle of $\mathbb{D}$ and $\mathbb{B}$
\begin{equation}
\begin{aligned}
 \text{bedding surface:}\quad\mathbb{S}=T^2\times\left\{\hf\right\}.
\end{aligned} 
\label{eq:def-S}
\end{equation}
Then, the cross section $D^2$ of the solid torus $D^2\times S^1$
is schematically depicted as
\begin{equation}
\begin{aligned}
\tikz[baseline={([yshift=-.5ex]current bounding box.center)},scale=1.6]{
\draw[fill,magenta,opacity=0.3] (0,0) circle (0.5);
\draw[thick] (0,0) circle (0.5);
\draw[thick] (0,0) circle (1);
\draw[thick] (0,0) circle (1.5);
\node at (0,0) {\text{core}};
\node at (0.46,0) [right] {$\mathbb{D}$};
\node at (0.96,0) [right] {$\mathbb{S}$};
\node at (1.46,0) [right] {$\mathbb{B}$};
}~~.
\end{aligned}
\label{eq:cross-section}
\end{equation}

\subsection{Core}
Let us first consider the ``core'' in \eqref{eq:core-sandwich}.
This part can be thought of as the state preparation for the 
defect $|\cD\ket$ in \eqref{eq:Z-D}.
Our interpretation is based on the relation between the
Reshetikhin-Turaev surgery invariant 
\cite{reshetikhin1991invariants} and the Turaev-Viro state sum
\cite{turaev1992state,Barrett:1993ab}.
As shown in \cite{kirillov2010,turaev2013}, two invariants 
for a 3-manifold $M$ are related by
\begin{equation}
\begin{aligned}
 \text{RT}_{\cZ(\cC)}(M)=\text{TV}_\cC(M),
\end{aligned} 
\label{eq:RT-TV}
\end{equation}
where $\cZ(\cC)$ denotes the Drinfeld center of $\cC$.
The Drinfeld center has a natural interpretation as 
the representation of the tube algebra
\begin{equation}
\begin{aligned}
 \cZ(\cC)=\text{Rep}(\text{Tube}(\cC)),
\end{aligned} 
\end{equation}
and the simple object $\mu\in\cZ(\cC)$
corresponds to an anyon in the Turaev-Viro TQFT $\text{TV}_\cC$.

When $\cC$ is a modular tensor category, 
the Drinfeld center $\cZ(\cC)$ becomes
\cite{izumi2000structure,mueger2001,izumi2001structure}
\begin{equation}
\begin{aligned}
 \cZ(\cC)=\b{\cC}\boxtimes\cC,
\end{aligned}
\label{eq:boxCC} 
\end{equation}
and the relation \eqref{eq:RT-TV} simplifies to
\begin{equation}
\begin{aligned}
 \bigl|\text{RT}_\cC(M)\bigr|^2=\text{TV}_\cC(M).
\end{aligned} 
\label{eq:RTsquare-TV}
\end{equation}
Note that the Reshetikhin-Turaev invariant can be thought of as a categorical
version of the Witten's Chern-Simons invariant \cite{Witten:1988hf}.
Then the relation \eqref{eq:RTsquare-TV} is analogous
to the Chern-Simons description of the $AdS_3$ gravity
\cite{Achucarro:1986uwr,Witten:1988hc}
\begin{equation}
\begin{aligned}
 S_\text{grav}=S_\text{CS}(A)+S_\text{CS}(\b{A}).
\end{aligned} 
\end{equation}

When the 3-manifold $M$ is a solid torus $D^2\times S^1$
in the ``core'',
its boundary is the topological boundary
$\mathbb{D}=\del(D^2\times S^1)$. 
Then the relation 
\eqref{eq:RT-TV} naturally gives rise to a
mapping between the insertion of the Wilson line $W_\mu$
on the left hand side of \eqref{eq:RT-TV} and the state
$|\cD_\mu\ket\in \cH_\cC(\mathbb{D})$ on  
the right hand side of \eqref{eq:RT-TV}.
Note that the state space $\cH_\cC(\mathbb{D})$ 
of Turaev-Viro TQFT is spanned by a linear
combination of the fusion diagrams of $\cC$ drawn on 
$\mathbb{D}=T^2$.
We stress that the state space $\cH_\cC(\mathbb{D})$
of Turaev-Viro TQFT 
should not be confused with the Hilbert space $\cH_\text{total}$
in \eqref{eq:H-tot}.

This mapping $W_\mu\to |\cD_\mu\ket$
is essentially given by the idempotent
$P^\mu$ in the tube algebra \eqref{eq:P-min}, where $\mu\in\cZ(\cC)$
corresponds to $(c,d)\in \b{\cC}\boxtimes\cC$
under the identification \eqref{eq:boxCC}.
By the relation $\text{RT}_{\cZ(\cC)}=\text{TV}_\cC$ in
\eqref{eq:RT-TV},
the insertion of the Wilson line $W_{\mu\in\cZ(\cC)}$ 
along the longitude of the solid torus
is encoded as a 
linear
combination of the fusion diagrams of $\cC$ on $\mathbb{D}=T^2$
\begin{equation}
\begin{aligned}
 W_{\mu}~\to~
|\cD_\mu\ket=\sum_{a,b,c\in\text{Irr}(\cC)}(W_\mu)_{c}^{ab}
\bigl|\cT^{c}_{ab}\bigr\ket\in\cH_{\cC}(\mathbb{D}),
\end{aligned} 
\label{eq:map-W-D}
\end{equation}
where $(W_\mu)_{c}^{ab}$ is the coefficient appearing in 
the idempotent $P^\mu$
\begin{equation}
\begin{aligned}
 P^\mu=\sum_{a,b,c\in\text{Irr}(\cC)}(W_\mu)_{c}^{ab}\cT^{c}_{ab},
\end{aligned} 
\end{equation}
which satisfies $P^{\mu}P^{\mu'}=\cob^{\mu\mu'}P^{\mu}$.
For the Ising model, the explicit value of the 
coefficient $(W_\mu)_{c}^{ab}$
can be read off from the idempotent $P^\mu=P^{(c,d)}$ in \eqref{eq:P-Ising}.
The mapping \eqref{eq:map-W-D}
is schematically depicted as
\begin{equation}
\begin{aligned}
\tikz[baseline={([yshift=-.5ex]current bounding box.center)},scale=1.6]{
\draw[thick] (0.5,1) arc (0:360: 0.5 and 0.2); 
\draw[thick] (-0.5,0) arc (180:360: 0.5 and 0.2); 
\draw[very thick,magenta] (0,0)--(0,1);
\draw[fill,magenta] (0,0) circle (0.03);
\draw[fill,magenta] (0,1) circle (0.03);
\draw[thick,dotted] (0.5,0) arc (0:180: 0.5 and 0.2); 
\draw[thick] (0.5,0) -- (0.5,1);
\draw[thick] (-0.5,0) -- (-0.5,1);
\node at (0,-0.2) [below] {};
\node at (0,0.4) [left] {$\mu$};
}\quad
=~
\sum_{a,b,c\in\text{Irr}(\cC)}(W_\mu)_{c}^{ab}~
\tikz[baseline={([yshift=-.5ex]current bounding box.center)},scale=1.6]{
\draw[thick] (0.5,1) arc (0:360: 0.5 and 0.2); 
\draw[thick] (-0.5,0) arc (180:360: 0.5 and 0.2); 
\draw[thick,dotted] (0.5,0) arc (0:180: 0.5 and 0.2); 
\draw[thick] (0.5,0) -- (0.5,1);
\draw[thick] (-0.5,0) -- (-0.5,1);
\draw[thick,blue] (0,-0.2) to [out=90, in=250] (0.5,0.4);
\draw[thick,blue] (-0.5,0.4) to [out=-70, in=270] (0,0.8);
\draw[thick,dashed,blue] (0.5,0.4) arc (0:180: 0.5 and 0.2); 
\draw[thick,blue] (0.2,0.1) -- (-0.2,0.45);
\node at (0.08,0.2) [above] {$c$};
\node at (0,-0.2) [below] {$a$};
\node at (0,0.75) [above] {$a$};
\node at (0.5,0.4) [right] {$b$};
\node at (-0.5,0.4) [left] {$b$};
}~,
\end{aligned} 
\label{eq:map-mu-D}
\end{equation}
where the top and the bottom of this figure are identified.
Namely, the Wilson line $W_\mu$ is encoded as a network of defects
on the boundary $T^2$ of the solid torus.
This mapping $W_\mu\to|\cD_\mu\ket$ can be thought of as the 
state preparation of $|\cD_\mu\ket\in\cH_\cC(\mathbb{D})$.
In other words, $|\cD_\mu\ket$
is a Hartle-Hawking state associated with the ``core'' cobordism 
$\emptyset\xrightarrow{\text{solid~torus}}\mathbb{D}$
in \eqref{eq:core-sandwich}.
\footnote{Our interpretation of the state preparation is slightly different
from \cite{Lin:2022dhv}.
In \cite{Lin:2022dhv}, a state preparation
of the physical boundary $|\mathbb{B}\ket$ was considered.
We are considering the 
state preparation of the topological boundary 
$|\mathbb{D}\ket=|\cD_\mu\ket$, not $|\mathbb{B}\ket$.
}

\subsection{Sandwich}
Next, let us consider the ``sandwich'' in \eqref{eq:core-sandwich}.
The Turaev-Viro TQFT on the cylinder 
$T^2\times I$
defines a homomorphism from the state space
$\cH_\cC(\mathbb{D})$ on the topological boundary to the
state space
$\cH_\cC(\mathbb{B})$ on the physical boundary  
\begin{equation}
\begin{aligned}
 \text{TV}_\cC(T^2\times I): ~\cH_\cC(\mathbb{D})
~\to~\cH_\cC(\mathbb{B}). 
\end{aligned} 
\label{eq:hom-T-T}
\end{equation}
Since the concatenation $(T^2\times I)\cup (T^2\times I)$
of two cylinders is again a cylinder $T^2\times I$, the homomorphism 
$\Pi=\text{TV}_\cC(T^2\times I)$ is an idempotent
\begin{equation}
\begin{aligned}
 \Pi^2=\Pi.
\end{aligned} 
\end{equation}
As shown in \cite{kirillov2011stringnetmo}, 
the image of $\Pi$ is isomorphic to the ground space
of the Levin-Wen string net model 
\begin{equation}
\begin{aligned}
 \text{Im}\,\Pi=\text{ker}\,H_\text{LW},
\end{aligned} 
\label{eq:kirillov}
\end{equation}
where the Hamiltonian $H_\text{LW}$ of the Levin-Wen model
is given by \cite{Levin:2004mi}
\begin{equation}
\begin{aligned}
 H_\text{LW}=\sum_v(1-A_v)+\sum_p(1-B_p).
\end{aligned} 
\end{equation}
Here $A_v$ and $B_p$ are commuting projectors defined at the vertex $v$
and the plaquette $p$, respectively. See \cite{Levin:2004mi}
for the details of the definition of $A_v$ and $B_p$.
As discussed in \cite{Okuyama:2026aoa},
the Levin-Wen Hamiltonian $H_\text{LW}$ generates a translation
along the radial direction $I$ and the
condition for the string-net state $|\Psi\ket$
\begin{equation}
\begin{aligned}
 H_\text{LW}|\Psi\ket=0
\end{aligned} 
\end{equation} 
can be thought of as the Wheeler-DeWitt equation in the radial
direction.
From \eqref{eq:kirillov},
the string-net state $|\Psi\ket$ also satisfies 
$\Pi|\Psi\ket=|\Psi\ket$.
In the Levin-Wen model, the fusion diagram $\cF$ is called the 
string-net.

The overlap $\bra\cD|\Pi|\cF\ket$ in \eqref{eq:Z-D}
is naturally interpreted as the Turaev-Viro state sum defined by
the homomorphism \eqref{eq:hom-T-T}, where $|\cD\ket$ and $|\cF\ket$
are states on the topological boundary and the physical boundary, respectively
\begin{equation}
\begin{aligned}
 |\cD\ket\in \cH_\cC(\mathbb{D}),\qquad
|\cF\ket\in \cH_\cC(\mathbb{B}).
\end{aligned} 
\end{equation}
As we discussed in the previous subsection, $|\cD\ket$ is obtained by the state
preparation in the core region.
Although $|\cD\ket$ and $|\cF\ket$ are both given by the fusion diagram
on $T^2$, their interpretations are different.
$|\cD\ket$ is a network of macroscopic defects, while
the fusion diagram $|\cF\ket$ is defined by the domain wall 
in the UV lattice model \eqref{eq:fusion-lattice}.
In other words, $|\cD\ket$ is a macroscopic state in the IR while
$|\cF\ket$ is a microscopic state in the UV.
This is consistent with the holographic picture that
the radial direction $I$ in the bulk corresponds to the energy scale
on the boundary theory
\begin{equation}
\begin{aligned}
  \underset{|\cD\ket\,=\,\text{IR}}{\mathbb{D}}
~\xrightarrow{~T^2\times I~}
~\underset{|\cF\ket\,=\,\text{UV}}{\mathbb{B}}~.
\end{aligned} 
\end{equation}

\subsection{Turaev-Viro state sum}
Let us consider the Turaev-Viro state sum 
description of the overlap $\bra\cD|\Pi|\cF\ket$ in 
\eqref{eq:Z-D}.
It turns out that the height variables are naturally
thought to be living on the bedding surface $\mathbb{S}$ \cite{Aasen:2020jwb}.
Following \cite{turaev2013,turaev2017monoidal}, 
we discretize the 3-manifold $T^2\times I$
into a 2-skeleton $P$
and assign 
the ``colors'' $c\in\text{Irr}(\cC)$
to the faces of the skeleton $P$.
\footnote{
In the literature, the color $c\in\text{Irr}(\cC)$
is sometimes assigned to the edges of the skeleton.
Assignments of the color to the faces or the edges
are related by the Poincar\'{e} duality in three 
dimensions. In this paper, 
following the approach
in \cite{turaev2013,turaev2017monoidal}
we consider the assignment of colors
to the faces.
}

For simplicity, let us first consider the case where
there is no defect on the topological boundary $\mathbb{D}$
\begin{equation}
\begin{aligned}
|0\ket\in\cH_\cC(\mathbb{D}). 
\end{aligned} 
\end{equation}
In this case, there is no non-trivial part of the 2-skeleton between
$\mathbb{D}$ and $\mathbb{S}$, i.e. the region between
$\mathbb{D}$ and $\mathbb{S}$ is empty.
The boundary of the 2-skeleton $P$ 
should agree with the fusion diagram $\cF$ on the 
physical boundary $\mathbb{B}$
\begin{equation}
\begin{aligned}
 \del P=\cF.
\end{aligned}
\label{eq:delP-F} 
\end{equation}
The coloring of $\cF$ naturally extends to the faces of $P$.
On the bedding surface $\mathbb{S}$, we draw the same fusion diagram
$\cF$ as the one on the physical boundary $\mathbb{B}$,
and we assign the height $a=0,1$ (or the color $\id,\psi$)
to the domains inside and outside of the domain wall.
We also assign color $\si$ to the disk inside the
green circle in \eqref{eq:fusion-lattice}.
A typical example of the fusion diagram $\cF$ on $\mathbb{B}$
and the coloring of $\mathbb{S}$ looks like
\begin{equation}
\begin{aligned}
 \tikz[baseline={([yshift=-.5ex]current bounding box.center)}]{
\coordinate (s0) at (0.5,0.5);
\coordinate (s1) at (1.5,0.5); 
\coordinate (s2) at (2.5,0.5); 
\coordinate (s3) at (3.5,0.5); 
\coordinate (t0) at (0.5,1.5);
\coordinate (t1) at (1.5,1.5); 
\coordinate (t2) at (2.5,1.5); 
\coordinate (t3) at (3.5,1.5); 
\coordinate (u0) at (0.5,2.5);
\coordinate (u1) at (1.5,2.5); 
\coordinate (u2) at (2.5,2.5); 
\coordinate (u3) at (3.5,2.5);
\draw[thick] (0,0)--(4,0)--(4,3)--(0,3)--(0,0);
\fill[red] (0.5,0.5)--(2.5,0.5)--(2.5,1.5)--(1.5,1.5)--(1.5,2.5)--(0.5,2.5)--cycle;
\draw[fill,green] 
(s0) circle (0.2) (s1) circle (0.2) (s2) circle (0.2) (s3) circle (0.2)
(t0) circle (0.2) (t1) circle (0.2) (t2) circle (0.2) (t3) circle (0.2)
(u0) circle (0.2) (u1) circle (0.2) (u2) circle (0.2) (u3) circle (0.2);
\node at (2,3) [above] {$\mathbb{S}$};
}
\hskip10mm
 \tikz[baseline={([yshift=-.5ex]current bounding box.center)}]{
\coordinate (s0) at (0.5,0.5);
\coordinate (s1) at (1.5,0.5); 
\coordinate (s2) at (2.5,0.5); 
\coordinate (s3) at (3.5,0.5); 
\coordinate (t0) at (0.5,1.5);
\coordinate (t1) at (1.5,1.5); 
\coordinate (t2) at (2.5,1.5); 
\coordinate (t3) at (3.5,1.5); 
\coordinate (u0) at (0.5,2.5);
\coordinate (u1) at (1.5,2.5); 
\coordinate (u2) at (2.5,2.5); 
\coordinate (u3) at (3.5,2.5);
\draw[thick] (0,0)--(4,0)--(4,3)--(0,3)--(0,0);
\draw[green,thick] 
(s0) circle (0.2) (s1) circle (0.2) (s2) circle (0.2) (s3) circle (0.2)
(t0) circle (0.2) (t1) circle (0.2) (t2) circle (0.2) (t3) circle (0.2)
(u0) circle (0.2) (u1) circle (0.2) (u2) circle (0.2) (u3) circle (0.2);
\draw[thick,red] (1.5,2.3)--(1.5,1.7);
\draw[thick,red] (1.7,1.5)--(2.3,1.5);
\draw[thick,red] (2.5,1.3)--(2.5,0.7);
\draw[thick,red] (2.3,0.5)--(1.7,0.5);
\draw[thick,red] (1.3,0.5)--(0.7,0.5);
\draw[thick,red] (0.5,2.3)--(0.5,1.7);
\draw[thick,red] (0.5,1.3)--(0.5,0.7);
\draw[thick,red] (1.3,2.5)--(0.7,2.5);
\node at (2,3) [above] {$\mathbb{B}$};
}
\end{aligned} 
\label{eq:S-conf1}
\end{equation}
Since we should sum over all possible configuration of heights
$a_v=0,1$ in \eqref{eq:Z-height}, we need to consider
the opposite coloring of \eqref{eq:S-conf1} 
on $\mathbb{S}$ as well
\begin{equation}
\begin{aligned}
 \tikz[baseline={([yshift=-.5ex]current bounding box.center)}]{
\coordinate (s0) at (0.5,0.5);
\coordinate (s1) at (1.5,0.5); 
\coordinate (s2) at (2.5,0.5); 
\coordinate (s3) at (3.5,0.5); 
\coordinate (t0) at (0.5,1.5);
\coordinate (t1) at (1.5,1.5); 
\coordinate (t2) at (2.5,1.5); 
\coordinate (t3) at (3.5,1.5); 
\coordinate (u0) at (0.5,2.5);
\coordinate (u1) at (1.5,2.5); 
\coordinate (u2) at (2.5,2.5); 
\coordinate (u3) at (3.5,2.5);
\fill[red] (0,0)--(4,0)--(4,3)--(0,3)--cycle;
\draw[thick] (0,0)--(4,0)--(4,3)--(0,3)--(0,0);
\fill[white] (0.5,0.5)--(2.5,0.5)--(2.5,1.5)--(1.5,1.5)--(1.5,2.5)--(0.5,2.5)--cycle;
\draw[fill,green] 
(s0) circle (0.2) (s1) circle (0.2) (s2) circle (0.2) (s3) circle (0.2)
(t0) circle (0.2) (t1) circle (0.2) (t2) circle (0.2) (t3) circle (0.2)
(u0) circle (0.2) (u1) circle (0.2) (u2) circle (0.2) (u3) circle (0.2);
\node at (2,3) [above] {$\mathbb{S}$};
}
\hskip10mm
 \tikz[baseline={([yshift=-.5ex]current bounding box.center)}]{
\coordinate (s0) at (0.5,0.5);
\coordinate (s1) at (1.5,0.5); 
\coordinate (s2) at (2.5,0.5); 
\coordinate (s3) at (3.5,0.5); 
\coordinate (t0) at (0.5,1.5);
\coordinate (t1) at (1.5,1.5); 
\coordinate (t2) at (2.5,1.5); 
\coordinate (t3) at (3.5,1.5); 
\coordinate (u0) at (0.5,2.5);
\coordinate (u1) at (1.5,2.5); 
\coordinate (u2) at (2.5,2.5); 
\coordinate (u3) at (3.5,2.5);
\draw[thick] (0,0)--(4,0)--(4,3)--(0,3)--(0,0);
\draw[green,thick] 
(s0) circle (0.2) (s1) circle (0.2) (s2) circle (0.2) (s3) circle (0.2)
(t0) circle (0.2) (t1) circle (0.2) (t2) circle (0.2) (t3) circle (0.2)
(u0) circle (0.2) (u1) circle (0.2) (u2) circle (0.2) (u3) circle (0.2);
\draw[thick,red] (1.5,2.3)--(1.5,1.7);
\draw[thick,red] (1.7,1.5)--(2.3,1.5);
\draw[thick,red] (2.5,1.3)--(2.5,0.7);
\draw[thick,red] (2.3,0.5)--(1.7,0.5);
\draw[thick,red] (1.3,0.5)--(0.7,0.5);
\draw[thick,red] (0.5,2.3)--(0.5,1.7);
\draw[thick,red] (0.5,1.3)--(0.5,0.7);
\draw[thick,red] (1.3,2.5)--(0.7,2.5);
\node at (2,3) [above] {$\mathbb{B}$};
}
\end{aligned} 
\label{eq:S-conf2}
\end{equation}
Then, the 2-skeleton $P$ is obtained by extending the fusion
diagram $\cF$ to the radial direction
and we assign the natural coloring $c(e)$ 
to the face $e\times I_{1/2}$ of $P$,
where $I_{1/2}=[1/2,1]$ is the interval between $\mathbb{S}$
and $\mathbb{B}$, $e$ is an edge of $\cF$, and $c(e)$
denotes the color of the edge $e$ in $\cF$. 
The 2-skeleton $P$ ends on the bedding surface $\mathbb{S}$ with the coloring of
the faces of $P\cap \mathbb{S}$ specified by the configuration 
of heights on $\mathbb{S}$
such as \eqref{eq:S-conf1} or \eqref{eq:S-conf2}.

Let us focus on the local structure near the 
trivalent junction $x$ on $\mathbb{S}$
\begin{equation}
\begin{aligned}
 \tikz[baseline={([yshift=-.5ex]current bounding box.center)}]{
\coordinate (s0) at (0,0);
\coordinate (s1) at (2,-1);
\coordinate (s2) at (2,2); 
\coordinate (s3) at (0,3);
\coordinate (sc) at (1,1);
\coordinate (sb) at (1,-0.5);
\coordinate (b0) at (3,0);
\coordinate (b1) at (5,-1);
\coordinate (b2) at (5,2);
\coordinate (b3) at (3,3);
\coordinate (bc) at (4,1);
\coordinate (bb) at (4,-0.5);
\draw[red,fill,opacity=0.2,very thick] (sb)--(bb)--(bc)--(sc)--cycle;
\draw[green,fill,opacity=0.2,very thick] (sc)--(bc)--(b2)--(s2)--cycle;
\draw[green,fill,opacity=0.2, very thick] (sc)--(bc)--(b3)--(s3)--cycle;
\draw[fill,green] (sc)--(s2)--(s3)--cycle;     
\draw[fill,red] (sc)--(s3)--(s0)--(sb)--cycle;       
\draw[thick] (s0)--(s1)--(s2)--(s3)--cycle;
\draw[thick] (b0)--(b1)--(b2)--(b3)--cycle;
\draw[very thick,red] (bb)--(bc);  
\draw[very thick,green] (bc)--(b2);  
\draw[very thick,green] (bc)--(b3);  
\draw[fill] (sc) ellipse (0.06 and 0.08);
\node at (1,-0.6) [below] {$\mathbb{S}$};
\node at (4,-0.6) [below] {$\mathbb{B}$};
\node at (sc) [right] {\,$x$};
}
\end{aligned}
\end{equation}
The weight $\text{Tet}(x)$ of the vertex $x$ is obtained as follows
\cite{turaev2013,turaev2017monoidal}:
we draw a small sphere $S^2_x$ around $x$
and consider 
the intersection of $S^2_x$ and the 2-skeleton $P$.
This intersection defines a fusion diagram on $S^2_x$, 
which is evaluated
as the $6j$-symbol denoted as $\text{Tet}(x)$
\begin{equation}
\begin{aligned}
\text{Tet}(x)~~=~~
 \tikz[baseline={([yshift=-.5ex]current bounding box.center)}]{
\draw[very thick,green,dashed] (2,0) to [out=160, in=-10] (-0.85,1);
\draw[thick] (0,-2) arc (-90:90: 1 and 2);
\draw[thick] (0,0) circle (2);
\draw[very thick,green] (0,2) arc (90:30: 1 and 2);
\draw[very thick,green,dashed] (0,2) arc (90:150: 1 and 2);
\draw[very thick,red,dashed] (0,-2) arc (270:150: 1 and 2);
\draw[very thick,red] (0,-2) arc (-90:0: 2 and 2);
\draw[very thick,green] (2,0) to [out=120, in=-20] (0.85,1);
\draw[fill] (0,0) ellipse (0.06 and 0.08);
\node at (0,-0.1) [below] {$x$};
\node at (-1,-0.3) [left] {$\psi$};
\node at (0.9,-0.3) [right] {$\id$};
\node at (1.6,-1.3) [right] {$\psi$};
\node at (0,2) [above] {$\si$};
\node at (0,0.8) [above] {$\si$};
\node at (1.4,0.75) [above] {$\si$};
}
~~=~~
 \tikz[baseline={([yshift=-.5ex]current bounding box.center)}]{
\draw[thick] (0,0) circle (1);
\draw[thick] (0,0)--(0,-1);
\draw[thick] (0,0)--(0.86,0.5); 
\draw[thick] (0,0)--(-0.86,0.5); 
\node at (0,1) [above] {$\si$};
\node at (0.4,0.22) [above] {$\si$};
\node at (-0.4,0.22) [above] {$\si$};
\node at (0.8,-0.6) [right] {$\id$};
\node at (-0.8,-0.6) [left] {$\psi$};
\node at (0,-0.5) [right] {$\psi$};
}
~~=\left[
\begin{matrix}
 \si~ &~\si~ &~\psi\\
\psi~&~\id~ &~\si
\end{matrix}
\right].
\end{aligned}
\end{equation}
This reproduces the Boltzmann weight in \eqref{eq:diamond}
up to a normalization factor of $d_\si$.
Note that the diamond region in \eqref{eq:diamond} contains
two trivalent junctions, and hence the 
Boltzmann weight in \eqref{eq:diamond} is given by the product of 
two $6j$-symbols.
Then, the Turaev-Viro state sum is written as
\begin{equation}
\begin{aligned}
 \bra 0|\Pi|\cF\ket=\sum_{\text{heights}}\prod_x \text{Tet}(x),
\end{aligned} 
\label{eq:TV-sum}
\end{equation}
where $x$ runs over all vertices of the 2-skeleton $P$.

For the configuration in \eqref{eq:S-conf1},
the 2-skeleton $P$ is schematically depicted as
\begin{equation}
\begin{aligned}
 \tikz[baseline={([yshift=-.5ex]current bounding box.center)}]{
\draw[thick] (0,0)--(2,-1)--(2,2)--(0,3)--cycle;
\draw[thick] (3,0)--(5,-1)--(5,2)--(3,3)--cycle;
\draw[very thick, green] (0.5,0.5)--(3.5,0.5);
\draw[very thick, green] (1.5,0)--(4.5,0);
\draw[very thick, green] (1.5,1)--(4.5,1);
\draw[very thick, green] (1,1.25)--(4,1.25);
\draw[very thick, green] (1,2)--(4,2);
\draw[very thick, green] (0.5,2.25)--(3.5,2.25);
\draw[thick,red,fill,opacity=0.2] (1.5,0)--(4.5,0)--(4.5,1)--(1.5,1)--cycle;
\draw[thick,red,fill,opacity=0.2] (1.5,1)--(4.5,1)--(4,1.25)--(1,1.25)--cycle;
\draw[thick,red,fill,opacity=0.2] (1.5,1)--(4.5,1)--(4,1.25)--(1,1.25)--cycle;
\draw[thick,red,fill,opacity=0.2] (1,1.25)--(4,1.25)--(4,2)--(1,2)--cycle;
\draw[thick,red,fill,opacity=0.2] (1,2)--(4,2)--(3.5,2.25)--(0.5,2.25)--cycle;
\draw[thick,red,fill,opacity=0.2] (0.5,2.25)--(3.5,2.25)--(3.5,0.5)--(0.5,0.5)--cycle;
\draw[thick,red,fill,opacity=0.2] (0.5,0.5)--(3.5,0.5)--(4.5,0)--(1.5,0)--cycle;
\draw[thick,red] (3.5,0.5)--(4.5,0)--(4.5,1)--(4,1.25)--(4,2)
--(3.5,2.25)--cycle;
\fill[red] (0.5,0.5)--(1.5,0)--(1.5,1)--(1,1.25)--(1,2)
--(0.5,2.25)--cycle;
\node at (0.5,2.75) [above]{$\mathbb{S}$};
\node at (3.5,2.75) [above]{$\mathbb{B}$};
\node at (4.4,0.4) [right]{$\mathcal{F}$};
\node at (2.5,0.5) {$P$};
}~.
\end{aligned}
\label{eq:P-bulk}
\end{equation}
Here we have suppressed the green circles for simplicity.
This picture suggests the following interpretation
\begin{equation}
\begin{aligned}
 \text{2-skeleton}~P~=~\text{worldsheet of string-net}~\cF.
\end{aligned} 
\end{equation}
Since the Turaev-Viro state sum \eqref{eq:TV-sum} 
is a topological invariant under the various moves
of skeletons (known as the Matveev-Piergallini moves \cite{turaev2013,turaev2017monoidal}),
the picture in \eqref{eq:P-bulk} can be deformed arbitrarily
unless the non-topological 
part $A(\cF)$ in \eqref{eq:Z-sumF} does not change.
Using this topological invariance, the 2-skeleton $P$
in \eqref{eq:P-bulk}
can be deformed (schematically) as
\begin{equation}
\begin{aligned}
 \tikz[baseline={([yshift=-.5ex]current bounding box.center)}]{
\draw[thick] (0,0)--(2,-1)--(2,2)--(0,3)--cycle;
\draw[thick,red, name path=A] (1,0.3) arc (-90:90: 0.3 and 0.8); 
\draw[thick,red,dotted] (1,0.3) arc (-90:-270: 0.3 and 0.8); 
\draw[thick,red,name path=B] (1,0.3) to [out=180, in=-180] (1,1.9);
\tikzfillbetween[of=A and B]{red, opacity=0.2};
\node at (1.27,0.9) [right]{$\mathcal{F}$};
\node at (0.5,2.75) [above]{$\mathbb{B}$};
}~.
\end{aligned} 
\label{eq:worldsheet}
\end{equation}
This further suggests the picture 
that the boundary of the ``worldsheet'' $P$
ends on the domain wall $\cF$ on $\mathbb{B}$
\begin{equation}
\begin{aligned}
 \del(\text{worldsheet})=\text{domain wall of Ising spins},
\end{aligned} 
\end{equation}
and the color of the worldsheet represents the spins 
inside the domain wall.
This picture \eqref{eq:worldsheet} is 
reminiscent of the holographic dual
of the Wilson loop in $\mathcal{N}=4$ SYM
\cite{Maldacena:1998im}.

When there is a defect $\cD$ on $\mathbb{D}$, 
there is a non-trivial 2-skeleton between $\mathbb{D}$
and $\mathbb{S}$,
and the condition \eqref{eq:delP-F} is generalized as
\begin{equation}
\begin{aligned}
 \del P=\cD\cup \cF.
\end{aligned} 
\end{equation} 
Then, the local
structure near a vertex $x$ on $\mathbb{S}$ looks like
\begin{equation}
\begin{aligned}
 \tikz[baseline={([yshift=-.5ex]current bounding box.center)}]{
\coordinate (s0) at (0,0);
\coordinate (s1) at (2,-1);
\coordinate (s2) at (2,2); 
\coordinate (s3) at (0,3);
\coordinate (sc) at (1,1);
\coordinate (b0) at (3,0);
\coordinate (b1) at (5,-1);
\coordinate (b2) at (5,2);
\coordinate (b3) at (3,3);
\coordinate (d0) at (-3,0);
\coordinate (d1) at (-1,-1);
\coordinate (d2) at (-1,2);
\coordinate (d3) at (-3,3);
\coordinate (x1) at (-3,1.5);
\coordinate (x2) at (-1,0.5);
\coordinate (x3) at (0,1.5);
\coordinate (x4) at (2,0.5);
\coordinate (a1) at (1,-0.5);
\coordinate (a2) at (1,2.5);
\coordinate (a3) at (4,2.5);
\coordinate (a4) at (4,-0.5);
\draw[thick] (s0)--(s1)--(s2)--(s3)--cycle;
\draw[thick] (b0)--(b1)--(b2)--(b3)--cycle;
\draw[thick] (d0)--(d1)--(d2)--(d3)--cycle;
\draw[fill,opacity=0.2] (x1)--(x2)--(x4)--(x3)--cycle;
\draw[blue,fill,opacity=0.2] (a1)--(a2)--(a3)--(a4)--cycle;
\draw[fill] (1,1) ellipse (0.06 and 0.08);
\node at (1,-0.6) [below] {$\mathbb{S}$};
\node at (4,-0.6) [below] {$\mathbb{B}$};
\node at (-2,-0.6) [below] {$\mathbb{D}$};
\node at (1,1) [below] {$x$};
\node at (-0.5,1) {$\al$};
\node at (2.5,1) {$\bt$};
\node at (1.5,1.5) {$h_1$};
\node at (0.5,2) {$h_2$};
\node at (1.5,0) {$h_4$};
\node at (0.5,0.5) {$h_3$};
}~,
\end{aligned}
\label{eq:int-D-B}
\end{equation}
where the gray face and the blue face are extended from the
defect on $\mathbb{D}$ and the domain wall on $\mathbb{B}$, 
respectively, and $h_1,\cdots,h_4$ are the height variables on
$\mathbb{S}$.
Again, we draw a small sphere
$S^2_x$
surrounding the vertex $x$
in \eqref{eq:int-D-B}.
By evaluating the fusion diagram
appearing from the intersection of $S^2_x$ and $P$, we 
obtain the weight $\text{Tet}(x)$ as 
\cite{turaev2013,turaev2017monoidal,Aasen:2020jwb}
\begin{equation}
\begin{aligned}
 \text{Tet}(x)=
\left[
\begin{matrix}
 \al~ &~h_1~ &~h_2\\
\bt~&~h_3~ &~h_4
\end{matrix}
\right].
\end{aligned} 
\end{equation}
Then the Turaev-Viro state sum 
in the presence of defect $\cD$ becomes
\begin{equation}
\begin{aligned}
 \bra\cD|\Pi|\cF\ket=\sum_{\text{heights}}\prod_x\text{Tet}(x).
\end{aligned} 
\end{equation}

\section{Conclusions and outlook}\label{sec:conclusion}
In this paper, we proposed a holographic dual of the 2d Ising model on
a torus as the Turaev-Viro TQFT on a solid torus.
The bulk spacetime is divided into the ``core'' region
and the ``sandwich'' region (see \eqref{eq:cross-section}
for the cross section of the solid torus).
The core prepares a defect state $|\cD\ket$
on the topological boundary $\mathbb{D}$.
The sandwich region computes the Turaev-Viro state sum
$\bra\cD|\Pi|\cF\ket$ between the state $|\cD\ket$ on 
the topological boundary $\mathbb{D}$ and the state $|\cF\ket$ on the physical
boundary $\mathbb{B}$.
The fusion diagram $\cF$ on the physical boundary
$\mathbb{B}$ is given by the AFM construction in 
\eqref{eq:fusion-lattice}.
There is also a bedding surface $\mathbb{S}$
between $\mathbb{D}$ and $\mathbb{B}$, and
the height variables are naturally living on the faces of
$\mathbb{S}$.
We discretize the bulk spacetime by the 
2-skeleton $P$ which interpolates
$\cD$ and $\cF$.  
When the defect on $\mathbb{D}$ is absent,
the 2-skeleton $P$ ends on the bedding surface
$\mathbb{S}$ and $P$ is interpreted as the bulk worldsheet of the 
string-net $\cF$ (see \eqref{eq:P-bulk} for 
a schematic picture).

There are many interesting open questions.
It would be interesting to generalize our approach to
models other than the Ising model.
In fact, the AFM construction can be applied to
the Andrew-Forester-Baxter model and the Potts model as well
\cite{Aasen:2020jwb} based on more complicated categories
than the Ising model.
It would be interesting to understand the bulk dual of such
models generalizing our approach in this paper.

The Ising model flows to the $c=1/2$ CFT in the IR.
Using the AdS/CFT correspondence,
the bulk dual of the Ising CFT has been studied
in \cite{Castro:2011zq,Jian:2019ubz,Romaidis:2023zpx,Janik:2025zji}.
Since the central charge of the Ising CFT is very small,
the bulk quantum gravity is 
in a highly quantum regime,
far away from the semi-classical gravity regime.
It would be interesting to consider a lattice model
which flows to a CFT 
with a large central charge.
One possibility is the symmetric product of 
$n$ Ising models considered in \cite{Dymarsky:2026lnf}.
This model has a central charge $c=n/2$ which becomes
large in the large $n$ limit.
It would be interesting to study the 
bulk dual of such a model along the lines of \cite{Aasen:2020jwb}
and generalizing our approach in this paper.

We considered the bulk dual of a lattice model
without taking the continuum limit, keeping the lattice
spacing finite.
It would be interesting to take a continuum
limit of our bulk construction and see if
it is related to the bulk 3d gravity
based on the Virasoro TQFT \cite{Collier:2023fwi,Collier:2024mgv,Hartman:2025cyj,Hartman:2025ula}
in the continuum limit (see also \cite{Chen:2024unp,Hung:2024gma,Bao:2024ixc}).

In our approach, the boundary theory is the Ising model
with a fixed coupling $J=J_c$ and
we do not take an ensemble average of the model.
It would be interesting to consider an 
ensemble average of the Ising model 
(e.g. a random-bond Ising model\footnote{See e.g. \cite{Dennis:2001nw}
and references therein.})
and see what happens to the bulk spacetime
after we take an ensemble average.
We leave these questions as interesting future problems.

\acknowledgments
The author is grateful to Yasuyuki Kawahigashi,
Kantaro Ohmori and Mengyang Zhang for helpful
discussions.
This work was supported
in part by JSPS KAKENHI 25K07300.

\appendix
\section{Sum over $SL(2,\mathbb{Z})$ images}\label{app:SL2}
In this appendix, we consider the sum over the
$SL(2,\mathbb{Z})$ images of the
vacuum character $|\chi_\id|^2$ of the Ising CFT.
It was observed in \cite{Castro:2011zq}
that this sum over $SL(2,\mathbb{Z})$ images reproduces 
$Z_\text{Ising}$ in \eqref{eq:Z-Isng-CFT}
\footnote{See also \cite{Jian:2019ubz}.}
\begin{equation}
\begin{aligned}
 Z_\text{vac}=|\chi_\id|^2~\leadsto~Z_\text{Ising}=|\chi_\id|^2+|\chi_\psi|^2+
|\chi_\si|^2.
\end{aligned} 
\end{equation}
We will reproduce this result from an explicit, elementary computation.

The vector $v(\tau)$ 
 of the characters of Ising CFT in \eqref{eq:v-tau}
transforms under $\ga\in SL(2,\mathbb{Z})$ as
\begin{equation}
\begin{aligned}
 v(\ga\tau)=R(\ga) v(\tau),
\end{aligned} 
\end{equation} 
where $R(\ga)$ on the right-hand side denotes the $3\times 3$ representation matrix
of $SL(2,\mathbb{Z})$, which is given by a combination of the matrix $S$
and $T$ in \eqref{eq:S-T}.
Under this transformation, $\cZ_M(\tau)$ in \eqref{eq:ZM} transforms as
\begin{equation}
\begin{aligned}
 \cZ_M(\ga\tau)=v^\dag(\tau)R(\ga)^\dag MR(\ga) v(\tau)=\cZ_{R(\ga)^\dag MR(\ga)}(\tau).
\end{aligned} 
\end{equation}
Note that 
the phase of $R(\ga)$ does not contribute to the transformation of 
$\cZ_M(\tau)$, i.e. 
we consider the projective representation of $SL(2,\mathbb{Z})$
\begin{equation}
\begin{aligned}
 R(\ga)\sim e^{\ri\th}R(\ga).
\end{aligned} 
\label{eq:upto-phase}
\end{equation}
Following \cite{Castro:2011zq}, we denote 
by $Z_\text{grav}$
the sum over the $SL(2,\mathbb{Z})$ 
images of $Z_\text{vac}=|\chi_\id|^2$.
As discussed in \cite{Castro:2011zq},
$Z_\text{grav}$ is written as a sum over the orbit of 
$\chi_\id$ under the $SL(2,\mathbb{Z})$ action
up to a phase \eqref{eq:upto-phase}
\begin{equation}
\begin{aligned}
 Z_\text{grav}=\sum_{\chi\in\text{Orb}(\chi_\id)}|\chi|^2.
\end{aligned} 
\label{eq:Zgrav-sum}
\end{equation}
It turns out that this orbit $\text{Orb}(\chi_\id)$
consists of 24 elements \cite{Castro:2011zq}.
To see this, it is convenient to introduce
the basis $\{\chi_\pm,\chi_\si\}$, where $\chi_\pm$ is defined as
\cite{Romaidis:2023zpx}
\begin{equation}
\begin{aligned}
 \chi_\pm=\frac{1}{\rt{2}}(\chi_\mathbbm{1}\pm\chi_\psi).
\end{aligned} 
\end{equation}
From \eqref{eq:S-T}, we can easily show that
\begin{equation}
\begin{aligned}
 S(\chi_+)&=\chi_+,\quad S(\chi_-)=\chi_\si,\quad S(\chi_\si)=\chi_{-},\\
T(\chi_\pm)&=\chi_\mp,\quad T(\chi_\si)=\th_\si \chi_\si.
\end{aligned} 
\label{eq:modular-v}
\end{equation}
Let us consider the orbit of $\chi_\id=\frac{\chi_{+}+\chi_{-}}{\rt{2}}$
under the action of $S$ and $T$.
Since $T(\chi_\id)=\chi_\id$, acting $T$ first does not generate a new element.
Thus we should act $S$ first on $\chi_\id$, and then act $T$ or $S$. 
We find that the orbit of $\chi_\id$ 
under $SL(2,\mathbb{Z})$ is given by
\begin{equation}
\begin{aligned}
\text{Orb}(\chi_\id)= \big\{\chi_m,S(\chi_m),T(\chi_m)\big|m=0,\cdots,7\,(\text{mod}\,8)\big\},
\end{aligned} 
\label{eq:orb-set}
\end{equation}
where
\begin{equation}
\begin{aligned}
\chi_m&=T^{2m}S(\chi_\mathbbm{1})=\frac{\chi_++\th_\si^{2m}\chi_\si}{\rt{2}},\\
S(\chi_m)&=ST^{2m}S(\chi_\mathbbm{1})= \frac{\chi_++\th_\si^{2m}\chi_-}{\rt{2}},\\
T(\chi_m)&=T^{2m+1}S(\chi_\mathbbm{1})=\frac{\chi_-+\th_\si^{2m+1}\chi_\si}{\rt{2}}.
\end{aligned} 
\end{equation}
Note that $S^2(\chi_m)=\chi_m,\,T^2(\chi_m)=\chi_{m+1}$ and
\begin{equation}
\begin{aligned}
 TS(\chi_m)&=\frac{\chi_{-}+\th_\si^{2m}\chi_{+}}{\rt{2}}
=\th_\si^{2m}\frac{\chi_{+}+\th_\si^{16-2m}\chi_{-}}{\rt{2}}=\th_\si^{2m}
S(\chi_{8-m}),\\
ST(\chi_m)&=
\frac{\chi_\si+\th_\si^{2m+1}\chi_{-}}{\rt{2}}
=\th_\si^{2m+1}
\frac{\chi_{-}+\th_\si^{15-2m}\chi_\si}{\rt{2}}
=\th_\si^{2m+1}T(\chi_{7-m}).
\end{aligned} 
\end{equation}
Thus, the set \eqref{eq:orb-set}
is closed under the action of $SL(2,\mathbb{Z})$
up to a $U(1)$ phase, and
hence we conclude that the $SL(2,\mathbb{Z})$ orbit of $\chi_\id$
is give by the 24 elements in \eqref{eq:orb-set}.
Then the sum in \eqref{eq:Zgrav-sum} becomes
\begin{equation}
\begin{aligned}
 Z_\text{grav}
&=\sum_{m=0}^7 \Bigl(|\chi_m|^2+
|S(\chi_m)|^2+
|T(\chi_m)|^2\Bigr)\\
&=\sum_{m=0}^7 \Biggl[|\chi_{+}|^2+|\chi_{-}|^2+|\chi_\si|^2\\
&\quad +\hf \th_\si^{2m}\bigl(\b{\chi}_{+}\chi_\si+\b{\chi}_{+}\chi_{-}
+\th_\si\b{\chi}_{-}\chi_\si\bigr)+\hf\b{\th}_\si^{2m}\bigl(\b{\chi}_\si\chi_{+}
+\b{\chi}_{-}\chi_{+}+\b{\th}_\si\b{\chi}_\si\chi_{-}\bigr)\Biggr].
\end{aligned} 
\end{equation}
Using the obvious relation
\begin{equation}
\begin{aligned}
 \sum_{m=0}^7 \th_\si^{2m}=
\sum_{m=0}^7 \b{\th}_\si^{2m}=0,
\end{aligned} 
\end{equation}
we find
\begin{equation}
\begin{aligned}
 Z_\text{grav}=8\bigl(|\chi_{+}|^2+|\chi_{-}|^2+|\chi_\si|^2\bigr)=8Z_\text{Ising}.
\end{aligned} 
\end{equation}
Namely, the sum over the $SL(2,\mathbb{Z})$ images of the vacuum character 
$|\chi_\id|^2$ reproduces the full partition function $Z_\text{Ising}$
up to a factor of $8$ \cite{Castro:2011zq}.
\bibliography{cite}
\bibliographystyle{utphys}

\end{document}